\documentclass[usenatbib]{mn2e}
\usepackage{mathrsfs,amsmath,amssymb,amstext}
\usepackage{lscape}
\usepackage{natbib}
\usepackage{epsfig}
\usepackage{color}
\usepackage{float}
\usepackage{graphicx}
\usepackage{gensymb}
\usepackage[justification=centering]{caption}
\usepackage[flushleft]{threeparttable}
\usepackage[breaklinks,colorlinks,citecolor=blue,linkcolor=magenta]{hyperref} 
\usepackage{footnote}
\makesavenoteenv{tabular}

\usepackage[all]{hypcap} 

\newcommand{\WISE}{\textsc{wise}}
\newcommand{\NEOWISE}{\textsc{neowise}}
\newcommand{\UNWISE}{un\textsc{wise}}
\newcommand{\GLIMPSE}{\textsc{glimpse}}

\begin{document}

\title[The Milky Way's X/P Structure]{Quantifying the (X/peanut)--shaped Structure of the Milky Way -- New Constraints on the Bar Geometry}
\author[Ciambur et al.]{Bogdan C. ~Ciambur$^1$\thanks{E-mail: bciambur@swin.edu.au},  Alister W. ~Graham$^1$, Joss Bland-Hawthorn$^2$ \\ \\
 $^1$Centre for Astrophysics and Supercomputing, Swinburne University of Technology, Hawthorn, VIC 3122, Australia \\
 $^2$Sydney Institute for Astronomy, School of Physics A28, University of Sydney, NSW 2006, Australia}

\maketitle

\begin{abstract}

The nature, size and orientation of the dominant structural components in the Milky Way's inner $\sim\!4$ kpc -- specifically the bulge and bar -- have been the subject of conflicting interpretations in the literature. We present a different approach to inferring the properties of the long bar which extends beyond the inner bulge, via the information encoded in the Galaxy's X/peanut (X/P)-shaped structure. We perform a quantitative analysis of the X/P feature seen in \WISE\ wide-field imaging at $3.4 \,\mu$m and $4.6 \,\mu$m. We measure the deviations of the isophotes from pure ellipses, and quantify the X/P structure via the radial profile of the Fourier $n=6$ harmonic (cosine term $B_6$). In addition to the vertical height and integrated `strength' of the X/P instability, we report an intrinsic radius of $R_{{\it \Pi},{\rm int}} = 1.67\pm0.27$ kpc, and an orientation angle of $\alpha = {37\degree}^{+ 7\degree}_{-10\degree}$ with respect to our line-of-sight to the Galactic Centre. Based on X/P structures observed in other galaxies,
we make three assumptions: (i) the peanut is intrinsically symmetric,
(ii) the peanut is aligned with the long Galactic bar, and (iii) their sizes are correlated. Thus the implication for the Galactic bar is that it is oriented at the same 37$\degree$ angle and has an expected radius of $\approx 4.2$ kpc, but possibly as low as $\approx$ 3.2 kpc. We  further investigate how the Milky Way's X/P structure compares with other analogues, and find that the Galaxy is broadly consistent with our recently established scaling relations, though with a moderately stronger peanut instability than expected. We additionally perform a photometric decomposition of the Milky Way's major axis surface brightness profile, accounting for spiral structure, and determine an average disc scale length of $h=2.54\pm 0.16$ kpc in the \WISE\ bands, in good agreement with the literature. 

\end{abstract}


\newcommand{\elli}{$\textsc{Ellipse}$}
\newcommand{\iso}{$\textsc{Isophote}$}
\newcommand{\bmo}{$\textsc{Bmodel}$}
\newcommand{\cmo}{$\textsc{Cmodel}$}
\newcommand{\ifit}{$\textsc{Isofit}$}
\newcommand{\rf}{\textcolor{red}{reference}}
\newcommand{\rad}{$R_{\rm maj}$}
\newcommand{\Rpiint}{$R_{{\it \Pi},{\rm int}}$}

\section{Introduction}\label{sec:Introduction}

Although the Sun's placement within the Galactic disc offers a restricted perspective of the Galaxy's central structural components, it has become generally accepted that the Milky Way is a barred galaxy (see \citealt{Gerhard2002} and \citealt{Merrifield2004} for reviews on the topic). Nevertheless, a consensus has yet to be reached on the exact details of its central components. There are conflicting interpretations in the literature with regard to the nature and geometry of the Galactic `bulge': whether it is a classical or pseudo-bulge or both, the primary bar or the inner part of a longer, thinner bar, etc. The notion of a long, thin bar extending beyond the triaxial `bulge' region ($10\degree < l < 30\degree$) was introduced by \cite{Hammersley+1994}, who found evidence for such a structure from star counts in the Galactic plane. Building upon this, \cite{Hammersley+2000}, \cite{Lopez_Corredoira+01, Lopez-Corredoira+2006} and \cite{Cabrera-Lavers+2007,Cabrera-Lavers+08} confirmed and characterised this long bar. Using red clump giant (RCG) stars -- which are approximate standard candles (\citealt{Stanek+1994}) -- as tracers of the bar's structure, they obtained a bar approximately 4 -- 4.5 kpc long and inclined at close to $\sim$ 43$\degree$ with respect to the Sun--(Galactic Centre) line-of-sight (see also \citealt{Sevenster+1999}). While other studies have reported lower bar viewing angles (38$\degree\pm6\degree$ in \citealt{Zasowski2012}; $30\degree\pm10\degree$ in \citealt{Francis&Anderson2012}), these results nevertheless point to a misalignment between the newly discovered long bar and the inner triaxial `bulge', which recent works place at an orientation angle of $\sim 20\degree$ -- $30\degree$ (\citealt{Babusiaux&Gilmore2005}, \citealt{Cao+2013}, \citealt{Wegg&Gerhard2013}).

The majority of barred galaxies display `boxy', or X/peanut (X/P)--shaped `bulges'. These structures occur when orbital resonances (\citealt{Combes+1990}) or buckling (\citealt{Raha+1991}) cause the bars' inner parts to thicken vertically and take the characteristic `X', or `peanut' shape when viewed in close to side-on (bar) and edge-on (disc) projection, while in face-on views they often take the form of a `bar-lens' (\citealt{Laurikainen+2011,Laurikainen+2014}; \citealt{Athanassoula+2015}, \citealt{Laurikainen&Salo2017}). Recently, \cite{Ciambur&Graham2016} (hereafter CG16) introduced a quantitative framework to characterise the properties of X/P structures, and additionally showed evidence, through a sample of twelve nearby galaxies with X/P `bulges', that peanuts obey specific scaling relations. As a typical barred spiral galaxy, the Milky Way's `bulge' too is X/P--shaped (\citealt{Weiland+94}, \citealt{Dwek+1995}, \citealt{Lopez-Corredoira+2005}, \citealt{Wegg&Gerhard2013}, \citealt{Ness&Lang2016}). Multiple studies of the distribution, chemistry and kinematics of the stellar populations in the `bulge' region support its X/P nature (e.g., \citealt{McWilliam&Zoccali2010}, \citealt{Ness+2012}, \citealt{Vasquez+2013}, \citealt{Zoccali+2014}, \citealt{Rojas-Arriagada+2014}, \citealt{Williams+2016}, \citealt{Joo+2017}), although see \cite{Lopez-Corredoira2016,Lopez-Corredoira2017} and \cite{Gran+2016}.   

From a dynamical point of view, the developing
picture asserts that the Milky Way's peanut and long bar are different parts of essentially the same structure, i.e., the X/P structure is the central, vertically thickened part of the long bar (\citealt{Combes+1990}, \citealt{Martinez-Valpuesta&Gerhard2011}, \citealt{Romero-Gomez+2011}, \citealt{Zoccali+2016}), despite the slight misalignment between the two components. In support of this scenario, \cite{WeggGerhard&Portail2015} appear to reconcile this misalignment and find a long bar angle between 28$\degree$ and 
33$\degree$, consistent with the orientation of the triaxial `bulge'.

Since X/P structures arise from, and are thus part of, galactic bars, one can infer information pertaining to the latter by studying the properties of the former. For the Milky Way in particular, the eastern and western hemispheres of the X/P structure, viewed as they are, at different distances relative to the Sun, contain ample information both in the radial (in-plane) and vertical (off-plane) directions with respect to the disc. This in principle can constrain the X/P structure's orientation, and by extension, that of the Galactic bar, relative to the Sun. Moreover, the radial extent of X/P structures in other galaxies appears to correlate well with the length of their associated bars, with recent studies placing the ratio $R_{X/P}/R_{\rm bar} \approx$ 0.4--0.5 (\citealt{LuettickeDettmar&Pohlen2000}, \citealt{Laurikainen&Salo2017}, \citealt{Erwin&Debattista2017}). Careful measurements of the Milky Way's X/P bulge therefore have the potential to reveal the geometry (extent and orientation) of the Galactic bar. This is one of the main goals of this study.

In this paper, we use for the first time the Milky Way's X/P structure as a proxy for the long bar, and thus constrain the latter's spatial extent and orientation angle based on the properties of the former. We characterise in detail the Milky Way's X/P feature and compare it with other nearby analogues. The remainder of the paper is structured as follows. \S \ref{sec:theory} provides a theoretical outline of the methodology employed to extract quantitative diagnostics of the peanut structure, based on \cite{Ciambur2015} (hereafter C15) and \cite{Ciambur&Graham2016}, as well as the peanut and bar geometric parameters. \S \ref{sec:analysis} presents the wide-field \WISE\ datasets and the analysis process, and the results are presented in \S \ref{sec:results}, where the Milky Way is also compared with other, local X/P galaxies. The results are interpreted and discussed in \S \ref{sec:discussion}, and finally we conclude with \S \ref{sec:conclusion}. Throughout this paper we employ Galactic co-ordinates and assume a distance of the Sun to the Galactic Centre of $R_0 = 8.2\pm0.1$ kpc (\citealt{Bland-Hawthorn&Gerhard2016}).

\section{Theory}\label{sec:theory}

C15 has suggested that X/P structures likely leave an imprint in the 6$^{\rm th}$ Fourier component of galaxy isophotes, specifically in the cosine term, $B_6$ (see Figure \ref{fig:b6_harmonic}). Subsequently, CG16 demonstrated with a sample of twelve known X/P galaxies that this is indeed the case, and further introduced a methodology for extracting quantitative peanut diagnostics from a galaxy's radial $B_6$ profile\footnote{The Fourier coefficients (including $B_6$) of a galaxy's isophotes vary with radius from the photocentre, such that each isophote has its own value. One can thus extract a radial $B_6$ profile.}.  

\subsection{The Quantitative X/P Parameters}\label{sec:xppar}

In this work we apply the CG16 methodology to extract the parameters of the Milky Way's X/P structure. We briefly summarise these diagnostics here, and refer the reader to the aforementioned papers for further details.

\begin{enumerate}

\item the peak value of the $B_6$ profile, denoted by ${\it \Pi}_{\rm max}$.

\item the projected X/P radius, or half-length ($R_{{\it \Pi}}$), corresponding to the (major axis) radius where ${\it \Pi}_{\rm max}$ occurs. Note that the true, intrinsic, radius of a peanut is generally only measurable from a galaxy image when the bar is viewed perfectly side-on, or when its viewing angle ($\alpha$ in our notation) is known. However, as we show in \S\ref{sec:geometry}, it is possible to directly constrain this angle for the special case of the Milky Way, due to our privileged location within the Galactic disc and relative proximity to the bar. Throughout the paper we denote the $intrinsic$ (deprojected) radius by $R_{{\it \Pi},{\rm int}}$, and employ the convention $\alpha = 0\degree$ for end-on, and $90\degree$ for side-on, orientation. 

\item the X/P height ($z_{{\it \Pi}}$) above the disc plane, a quantity computed from the isophote where ${\it \Pi}_{\rm max}$ occurs. In general this value depends on the disc's inclination with respect to the line of sight, reaching a maximum when the disc is edge-on. Fortunately, this is the case for the Milky Way, as the Sun is located roughly in the disc's plane with a planar offset of $z_0 = 25\pm5$ pc (\citealt{Juric+2008}).

\item the integrated X/P strength ($S_{\it \Pi}$) defined as:

\begin{equation}
\label{equ:sharpness}
S_{\it \Pi} = 100 \times \int_{R_1}^{R_2} B_6(R) dR \,,
\end{equation}

\noindent where the limits $R_1$ and $R_2$ enclose the part of the $B_6(R)$ profile above the peak's half-maximum (${\it \Pi}_{\rm max}/2$), and

\item the $B_6$ profile's width ($W_{\it \Pi}$), equal to the full width at half-maximum (i.e. $R_2 - R_1$). \\

\end{enumerate}

\begin{figure}
	\centering
	\includegraphics[width=1.02\columnwidth]{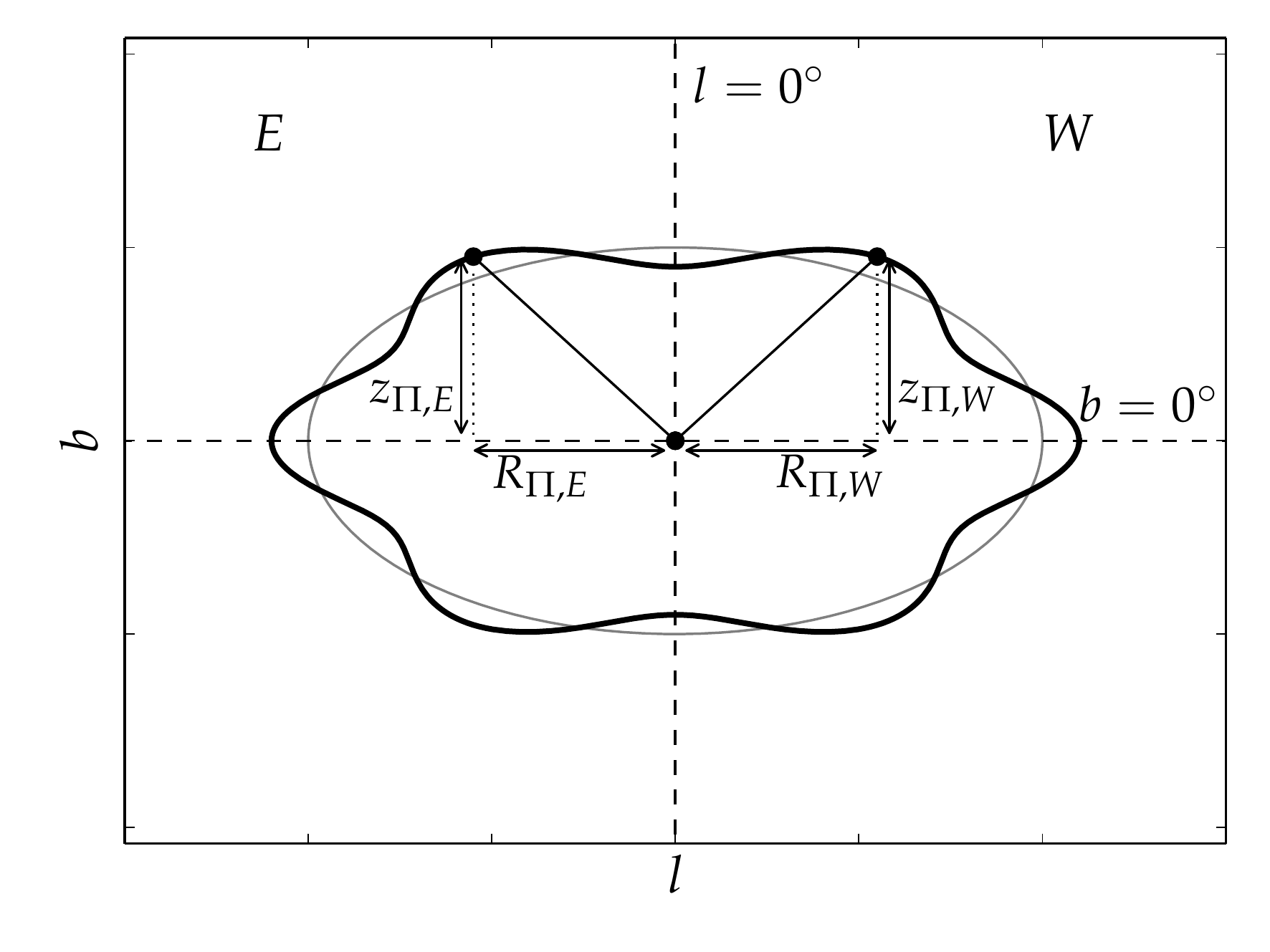}
	\caption{An X/P--shaped isophote (thick black), obtained by distorting an ellipse (thin grey) via a $n=6$ order Fourier harmonic (cosine term, $B_6=0.1$). The X/P projected radius ($R_{\it \Pi}$) and vertical height ($z_{\it \Pi}$) above the disc plane (i.e., the $b=0\degree$ plane) are derived from the isophote, as shown. Unlike the symmetric (side-on) X/P isophote shown above, the orientation angle and proximity of the Milky Way's X/P structure relative to the Sun induce an asymmetry in its isophotes about the $l=0\degree$ axis, such that the near ($East$) side appears larger, in projection, than the far ($West$) side, i.e., $R_{\it \Pi, E}>R_{\it \Pi, W}$ and $z_{\it \Pi, E}>z_{\it \Pi, W}$ (see also Figure \ref{fig:geometry}).}
	\label{fig:b6_harmonic}
\end{figure}

The galaxy isophote with the strongest $B_6$ perturbation, i.e., the isophote with semi-major axis associated with the peak of the radial $B_6$ profile (${\it \Pi}_{\rm max}$), defines the X/P structure's projected radius ($R_{\it \Pi}$) and height ($z_{\it \Pi}$) above the disc, as shown in Figure \ref{fig:b6_harmonic}. Note however that Figure \ref{fig:b6_harmonic} shows an X/P--shaped isophote that is symmetric about the $l=0\degree$ direction, as it would be observed in an external, edge-on galaxy with its bar oriented perpendicular to the line-of-sight. Our perspective of the Milky Way's X/P structure is from within the disc plane ($b=0\degree$), at relatively close proximity, and it is oriented at an angle with respect to the Sun--(Galactic Centre) line-of-sight, as illustrated in Figure \ref{fig:geometry}. This perspective induces an asymmetry in its isophotes, such that the near ($East$) `half' appears larger, in projection, than the far ($West$) `half', i.e., $R_{\it \Pi, E}>R_{\it \Pi, W}$ and $z_{\it \Pi, E}>z_{\it \Pi, W}$. This asymmetry warrants a separate treatment of the eastern and western hemispheres of our data, but offers the possibility to recover the $intrinsic$ radius and viewing angle of the X/P structure, as we show in the following subsection.

\subsection{The Geometry of the Problem}\label{sec:geometry}

\begin{figure}
\includegraphics[width=1.\columnwidth]{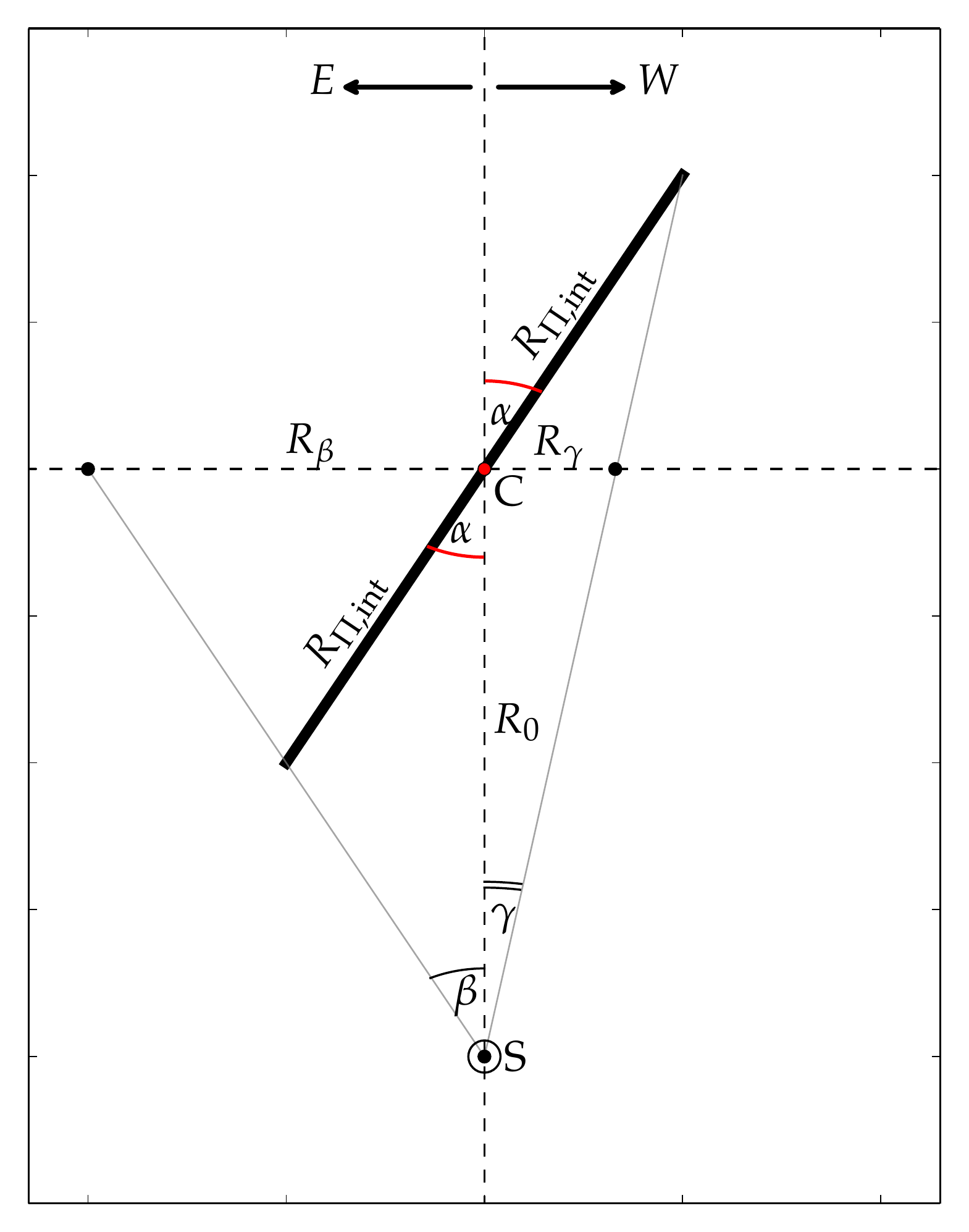}
\caption{A schematic representation of the (Sun+peanut) configuration, viewed from above the Galaxy. S represents the Sun, C the Galactic Centre, and their separation is denoted by $R_0$. The thick line represents the peanut structure, which has an intrinsic radius of \Rpiint, and makes an angle $\alpha$ with the line-of-sight from S to C. Finally, the projected angular sizes of the peanut, to the left ($E$) and to the right ($W$) of C, are labelled as $\beta$ and $\gamma$ respectively, and correspond to the projected radii $R_{\beta}$ and $R_{\gamma}$ at a distance $R_0$.}
\label{fig:geometry}
\end{figure}

The geometry of the (Sun -- peanut) configuration is illustrated schematically  in Figure \ref{fig:geometry}, and shows how the two `halves' of the peanut\footnote{This schematic holds for any symmetrically elongated structure viewed at relatively close proximity, such as the Galactic bar itself.}, which is oriented at an angle $\alpha$ with respect to our line-of-sight to the Galactic Centre (C), have different projected angular sizes. The half nearer to the Sun ($East$ of the Galactic Centre) has a larger angular size ($\beta$) while the more distant half ($West$ of the Galactic Centre) appears shorter ($\gamma$). The angles $\beta$ and $\gamma$, and the distance between the Sun and the Galactic Centre (i.e., SC $\equiv R_0$) are the only quantities needed to obtain the intrinsic (not apparent) radial extent of the peanut (\Rpiint) and orientation angle ($\alpha$), which are given by:

\begin{equation}\label{equ:peanut-length}
R_{{\it \Pi},{\rm int}} = \sqrt{R_{\beta}^2 (1-\eta) + R_0^2 \eta \left[ 1 - \frac{(1 - \eta)}{{\rm cos}^2(\beta)} \right]} ,
\end{equation}

\noindent where $R_{\beta}$ is the projected radius of the peanut eastward of C, on a plane located at a distance $R_0$ from the Sun, i.e., $R_{\beta} \equiv R_{\it \Pi,E} = R_0\: {\rm tan}(\beta)$, and $\eta$ is given by the ratio:

\begin{equation}\label{equ:eta}
\eta = \frac{R_{\beta} - R_{\gamma}}{R_{\beta} + R_{\gamma}} ,
\end{equation} 

\noindent where $R_{\gamma} (\equiv R_{\it \Pi,W})$ is the analogue of $R_{\beta}$, but westward of C (see Figure \ref{fig:geometry}). The orientation of the peanut structure, i.e., the angle $\alpha$ between the peanut and the line-of-sight towards the Galactic Centre, is given by:

\begin{equation}\label{equ:delta}
\alpha = {\rm cos}^{-1}\!\left(\eta \frac{R_0}{R_{\it \Pi}}\right) .
\end{equation}

The derivation of these equations, based on Stweart's theorem, is provided in Appendix \ref{sec:deriv}. Note that this framework operates on the assumption that the X/P structure is essentially 1D, as in Figure \ref{fig:geometry}. However, the bulge is by all accounts triaxial (\citealt{Perez-Villegas+2017}), and so its in-plane width, coupled with our perspective of it, adds some uncertainty. For example, in their Fig. 6, \cite{Lopez-Corredoira+2006} illustrate how the inclination angle of a triaxial ellipsoid viewed in projection can be over-estimated and, respectively, its intrinsic radius under-estimated, due to the different angular positions of the structure's true, and apparent (projected), ends. This effect is proportional to the in-plane `thickness' of the elongated structure, and to its length relative to $R_0$.


\section{Data Analysis}\label{sec:analysis}

\subsection{WISE Data}

To measure the properties of the Milky Way's X/P structure, we use two wide-field, infrared images (at 3.4 and 4.6\,$\mu$m) of the Galaxy, observed with the Wide-field Infrared Survey Explorer (\WISE) satellite (\citealt{Wright+2010}, \citealt{Mainzer+2014}). The images are identical to those used in \cite{Ness&Lang2016} except that they cover a slightly wider field of view. They were generated (D. Lang, private communication) by resampling the publicly released \NEOWISE-$Reactivation$\footnote{\url{http://neowise.ipac.caltech.edu/}} first-year data, particularly the ``\UNWISE" (\citealt{Lang2014}) co-adds from \cite{Meisner+2017}, into a Galactic coordinate system. 

One advantage of this particular dataset is that both images were observed in a wavelength regime where dust effects -- obscuration at shorter wavelengths and dust glow at longer -- are minimal, though still present (we discuss this further in \S \ref{sec:disc}). This can be readily noticed in Figure \ref{fig:wise}, which shows the raw $3.4\,\mu$m image (panel $a$) and $4.6\,\mu$m image (panel $c$). Moreover, performing our analysis on distinct datasets is useful for checking the robustness of the method, and results, to various biasing aspects, like data quality, or the amount/type of contamination (such as dust obscuration or extended bright sources, e.g., star clusters), which do not affect the two images the same.

\subsection{Pre-processing the Raw WISE Images}

\begin{figure*}
	\centering
	\includegraphics[width=0.8\textwidth]{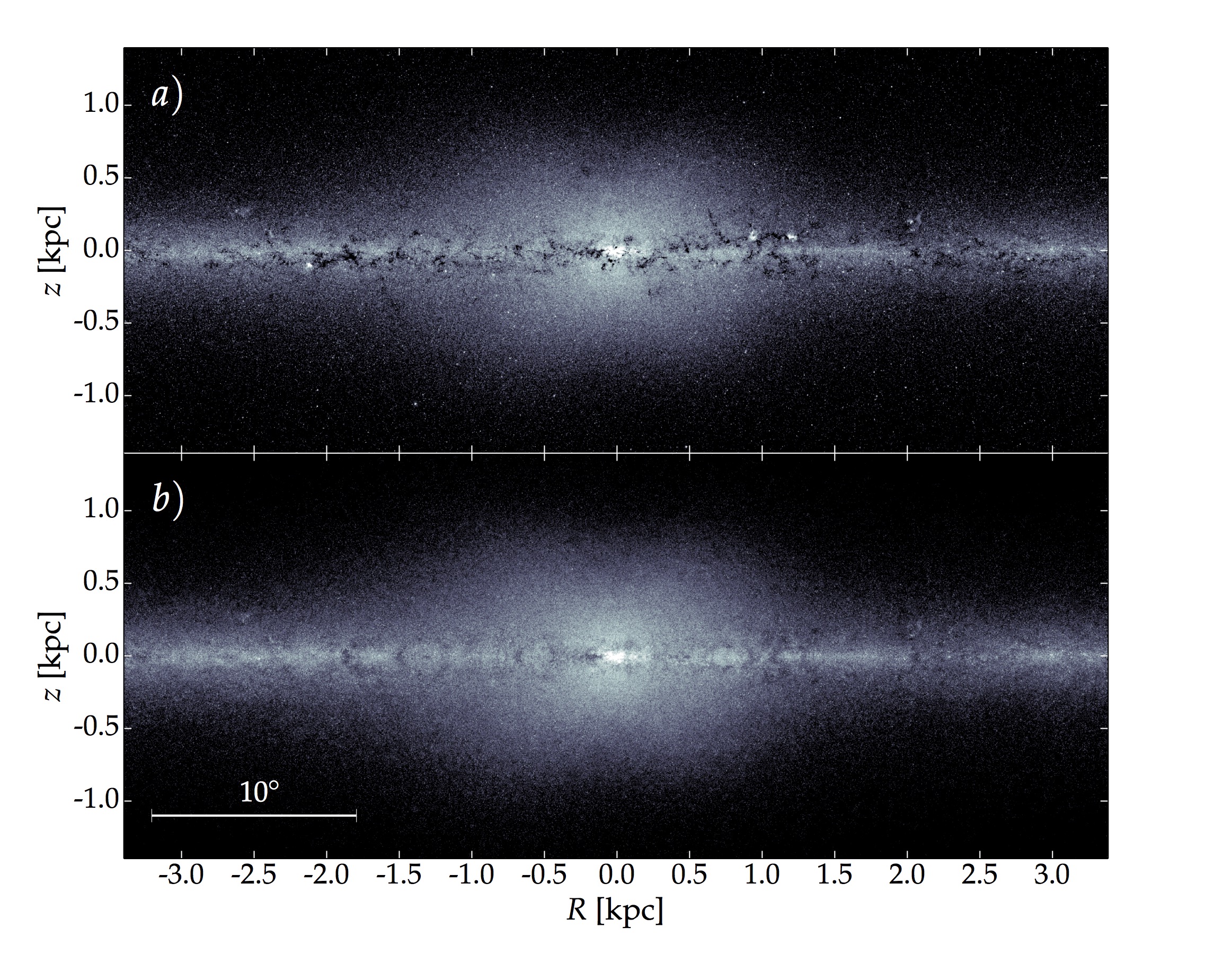}\\
	\includegraphics[width=0.8\textwidth]{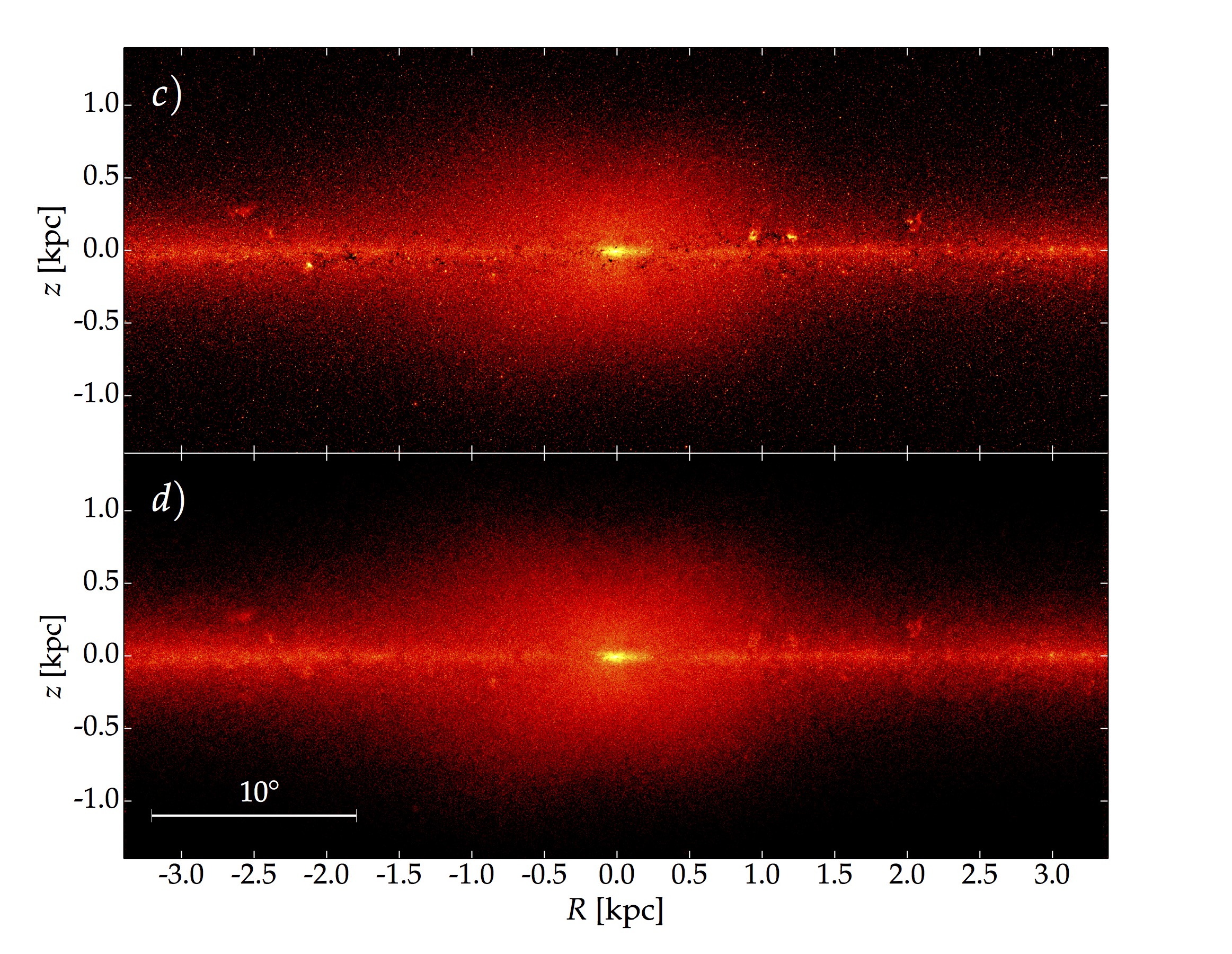}
	\caption{The Milky Way's X/peanut--shaped structure, observed by \WISE\ at $3.4\,\mu$m ($a$) and $4.6\,\mu$m ($c$). Scale assumes $R_0=8.2$ kpc. Image stretch adjusted to highlight the X/P structure. Panels $b)$ and $d)$ correspond to the results of our pre-processing by symmetric replacement process (see text) intended to reduce contamination from dust or extended sources like star clusters.}
	\label{fig:wise}
\end{figure*}

Before extracting the X/P parameters, both images were pre-processed in order to reduce, as much as possible, contamination from dust or bright sources such as star clusters, both visible in the raw images (Figure \ref{fig:wise}). This was done by taking advantage of the fact that such contamination is unlikely to occur symmetrically at both positive and negative Galactic latitudes ($b$ and $-b$), i.e, above and below the mid-plane, for a given Galactic longitude $l$. Each image was traversed pixel by pixel and, wherever a pixel of co-ordinates ($l,b$) was determined to have a value significantly offset from its local background (2.5$\sigma$ above or $2\sigma$ below the median within a 15$\times$15 pixel box around the pixel of interest), it was replaced by its symmetric counterpart ($l,-b$) on the opposite side of the disc mid-plane, provided that the latter pixel was not offset from its local background as well. The results of this pre-processing are displayed in Figure \ref{fig:wise}, panel $b)$ for the $3.4\,\mu$m observation and panel $d)$ for the $4.5\,\mu$m image. The pre-processed images were tested against the raw images by performing the subsequent analysis on both sets, and no systematic effect of the pre-processing was found. The various radial profiles extracted from the images (surface brightness profiles, ellipticity and $B_6$ profiles, etc.) did not differ in shape nor amplitude but only in the noise level, which was noticeably higher in the raw data.

The noise-reduced images were then convolved with a Gaussian kernel to produce a smoother (more diffuse) light distribution. This was done because \ifit, like most isophote-fitting codes, was designed to model external galaxies where the light is not discretised (individual stars are not resolved). Several values for the kernel size (dispersion $\sigma$) were tested and the value of $\sigma=5$ pixels was adopted, as it presented the best compromise between undersmoothing (light still discretised) and oversmoothing (erasing structures). 

Our relatively close proximity to the bar+peanut gives rise to an apparently asymmetric X/P structure, with a larger limb to the $East$ of the Galactic Centre and a smaller one to the $West$, as discussed in \S \ref{sec:geometry} (see also Figure \ref{fig:wise}). Consequently, the eastward and westward sides were modelled separately, in both images, by generating mirrored images reflected about the $l=0\degree$ axis. We show these four reflected images in Figure \ref{fig:reflections}, where panels $a$ and $b$ correspond to the near ($E$) and far ($W$) side reflections, respectively, for the 3.4\,$\mu$m data, while panels $c$ and $d$ are analogous, but for the 4.6\,$\mu$m data. Interestingly, panels $a$ and $c$ (the reflected near-side of the peanut, at both wavelengths) appear to display a slight additional asymmetry, between the northern and southern hemispheres of the X/P structure. In particular the `arms' of the X--shape seem to extend further apart at positive latitudes compared to negative latitudes. However, this apparent asymmetry is not evident in the reflected far-side images (panels $b$ and $d$).

\begin{figure*}
	\centering
	\includegraphics[width=0.49\textwidth]{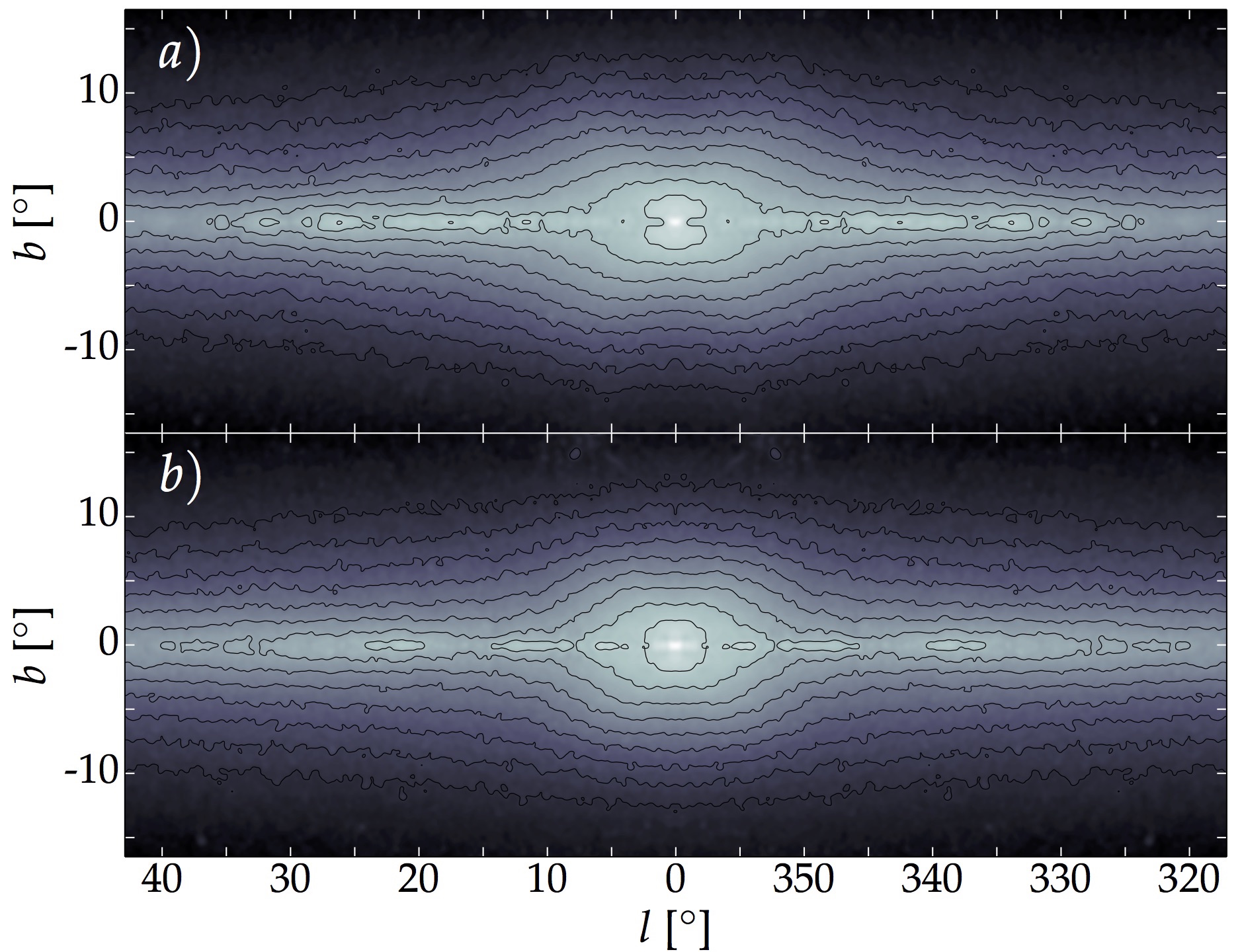}
	\includegraphics[width=0.49\textwidth]{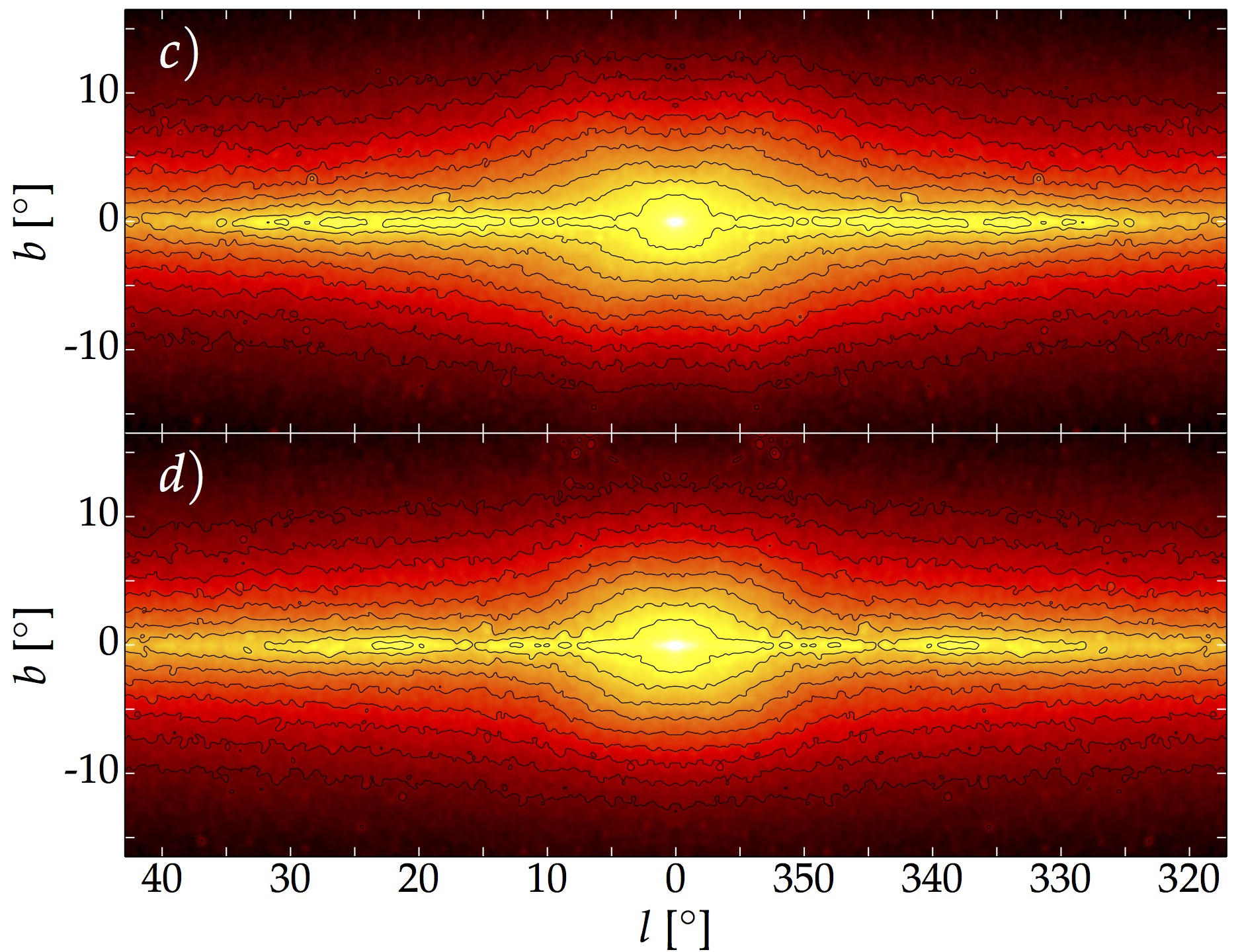}
	\caption{Milky Way images reflected about the $l=0\degree$ axis. \textbf{Left:} $3.4\,\mu$m image: $E$ hemisphere reflected to the $W$ ($a$) and vice versa ($b$). \textbf{Right:} $4.6\,\mu$m image: $E$ hemisphere reflected to the $W$ ($c$) and vice versa ($d$). Compared to Figure \ref{fig:wise}, the panels have a larger field-of-view, and the stretch has been adjusted to display a broader dynamical range. The contours are in 0.5 mag steps and the levels are the same in all four panels.}
	\label{fig:reflections}
\end{figure*}

The final step in preparing the data was to manually mask the four reflected images. In addition to the left-over regions still affected by dust (mostly at $3.4\,\mu$m), the (thin) disc was also masked. While CG16 retained the galaxy discs in their analysis (their 12 galaxies were also oriented nearly edge-on), the situation is different for the Milky Way because we are inside the disc. As such, the radial light profile along the mid-plane appears shallower than it would, were we observing from well outside the disc (i.e., the disc appears comparatively brighter at increasing distance from the centre than it would, were we not observing from within it). In order to avoid any biasing of the isophote shape caused by this effect, we thus excluded the major axis (the range $b= \pm \approx 2\degree\!.5$) and relied on the data in the remaining azimuthal range of the isophotes to constrain their shape. This effect is not important for the structural components of interest (bar, peanut) since the Sun is well outside of them. Manually masking the dust-affected regions is common practice in galaxy photometric modelling, and the results are usually robust to the amount of masking (except in extreme cases). This, coupled with the low levels of dust in our data (almost exclusively in the thin disc plane, which was already excluded for different reasons), did not warrant a more in-depth treatment of dust for this stage of the analysis.

\subsection{Modelling the Milky Way's X/P Structure}\label{sec:modellingX}

The image analysis was performed by running the isophote-fitting task \ifit~(C15). We ran \ifit~on the four processed images ($E$ and $W$ reflections, $3.4$ and $4.6\, \mu$m, Figure \ref{fig:reflections}), choosing a linear radial sampling step, fixing the isophotes' centre and position angle and allowing the ellipticity to vary.

The four resulting radial $B_6$ profiles are shown in Figure \ref{fig:b6_EW}. One can immediately discern the apparent asymmetry in the $B_6$ profile about the Galactic Centre ($l=0\degree$), caused by our perspective of the bar and peanut structure, as discussed in \S \ref{sec:geometry}. The two peaks where the peanut structure is a maximum, indicated by the vertical dashed lines in Figure \ref{fig:b6_EW}, mark the projected angular sizes of the two peanut limbs, which were computed to be: $\beta = 8\degree\!.25\pm0\degree\!.45$ and $\gamma = 5\degree\!.96\pm0\degree\!.44$. This same methodology for quantifying peanut sizes was employed in CG16. The full range in which the $B_6$ term is present in the isophotes extends roughly twice as far out ($\approx 16\degree\!.5\, W, -10\degree\!.5\, E$), at which point both sides curiously display a small `bump' just before reaching zero. The outer limits of positive $B_6$ are not of interest for our purposes, however, for several reasons. First, the outer `edge' of the $B_6$ signature corresponds to its faint outskirts, where the precise termination point of the feature becomes ambiguous due to noise -- this is seen in Figure \ref{fig:b6_EW} --  or to other photometric components, such as the disc, beginning to dominate the light (the disc is particularly relevant for the Milky Way, since we observe the X/P structure through the disc). Second, previous studies that have measured X/P structures relied on identification techniques (e.g., visual inspection, unsharp masking) that are sensitive to the point where the feature is strongest, not weakest. To keep consistency with the literature, on which we will draw in the following Sections, we remain within the CG16 framework and use the $B_6$ profile peak as the most reliable scale of the X/P structure. Nevertheless, the full range of the $B_6$ profile is still of interest, as it provides the width ($W_{\it \Pi}$) and `shape' of the profile, which are additional quantitative and, respectively, qualitative measures of peanut structure. Also apparent from Figure \ref{fig:b6_EW} is that the X/P structure is slightly more prominent in the redder $4.6\,\mu$m band than at $3.4\,\mu$m.

\begin{figure}
	\centering
	\includegraphics[width=1.02\columnwidth]{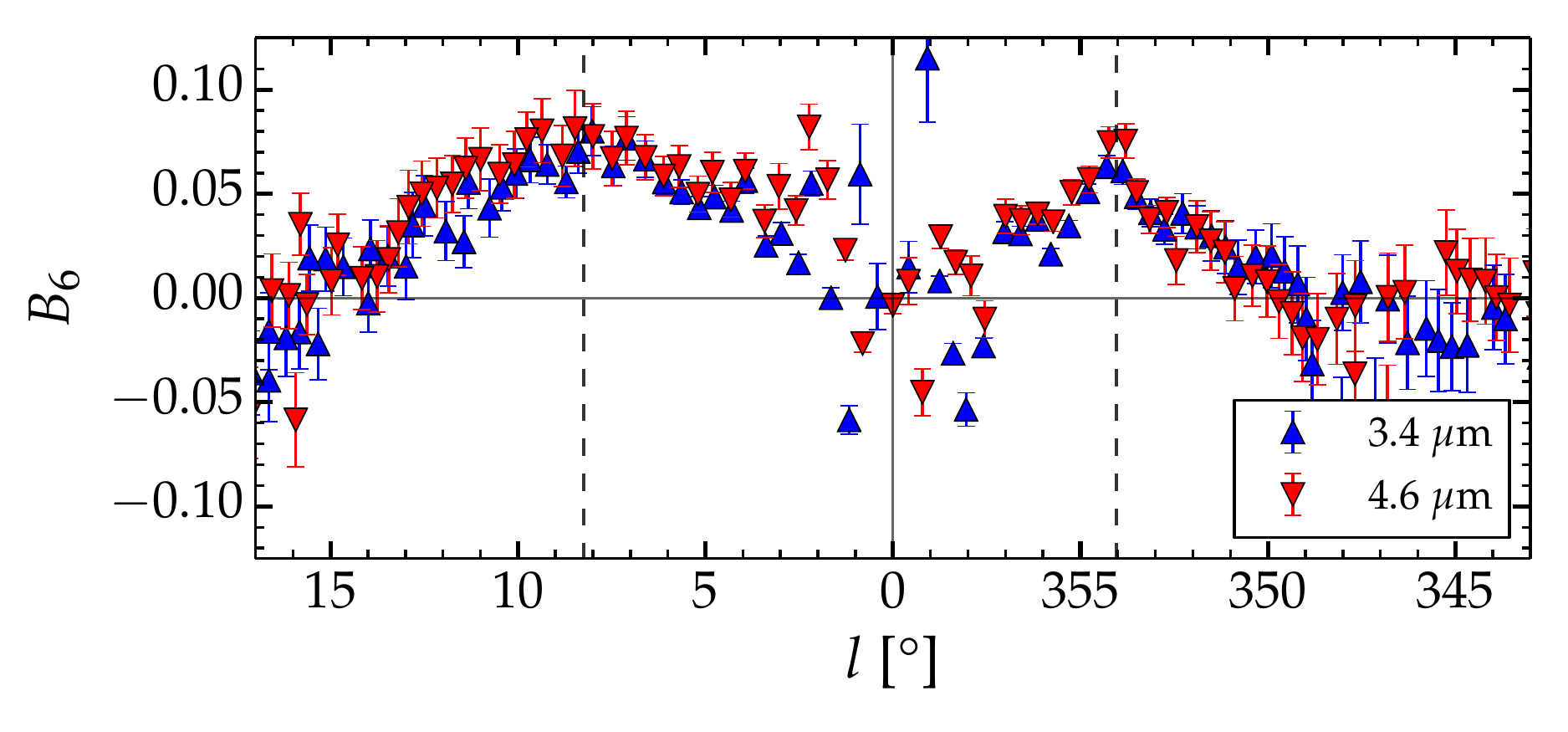}
	\caption{The $B_6$ harmonic amplitude as a function of Galactic longitude $l$. The $E$ and $W$ profiles peak at different projected angular distances ($\beta$ and $\gamma$ in Figure \ref{fig:geometry}) from the Galactic Centre due to our perspective of the Milky Way's bar/peanut structure. The locations of the two peaks, indicated by vertical dashed lines, allow for the computation of the length and viewing angle of the X/P structure and, by proxy, of the bar.}
	\label{fig:b6_EW}
\end{figure}


\section{Results}\label{sec:results}

\subsection{The (X/P Structure + Bar) Geometry}\label{sec:results_XPBarGeometry}

In \S \ref{sec:analysis} we have measured the apparent (projected) extent of the Milky Way's X/P structure, $E$ and $W$ of the Galactic Centre, which we shall now use to obtain the intrinsic radius of the peanut (\Rpiint) as well as its orientation angle $\alpha$ with respect to our line-of-sight to the centre of the Galaxy. We have determined the radial location of the $B_6$ profile peak in the two directions (Figure \ref{fig:b6_EW}) to be $\beta = 8\degree\!.25\pm0\degree\!.45$ and $\gamma = 5\degree\!.96\pm0\degree\!.44$. These yield an intrinsic radius of the X/P structure of $R_{{\it \Pi},{\rm int}} = 1.67\pm0.27$ kpc from Equation \ref{equ:peanut-length}, and an orientation angle of $\alpha = {37\degree}^{+ 7\degree}_{-10\degree}$ from Equation \ref{equ:delta}. The uncertainties have been computed according to Appendix \ref{sec:deriv}, using Equations \ref{equ:err_radius} ($\delta R_{\it \Pi}$) and \ref{equ:err_delta} ($\delta^{+,-}\!\alpha$). The outer bounds (east and west) where the $B_6$ profile declines to zero (see Figure \ref{fig:b6_EW}) could, in principle, also be used to constrain $\alpha$. Estimating these points to occur at $\approx 16\degree\!.5\, W, -10\degree\!.5\, E$ yields a value for the orientation angle of ${44\degree}^{+ 10\degree}_{-13\degree}$. However, as explained in \S \ref{sec:modellingX}, the greater statistical and systematic uncertainties, as well as possible biasing from disc light, associated with these outer radial locations make this measurement less reliable than using the $B_6$ peak, which we do throughout the analysis.

Multiple studies, based on stellar populations and numerical simulations, have shown evidence that the Milky Way's central `bulge' is not (primarily) the remnant of past merger events, i.e., a `classical' bulge, but rather it was built predominantly from disc stars through the buckling and secular evolution of the Galactic bar, the latter itself originating from the disc (\citealt{Shen+2010}, \citealt{Ness+2012,Ness+2013}; \citealt{DiMatteo+2014,DiMatteo2016}; \citealt{Abbott+2017}; see also \citealt{Fragkoudi+2017}). This result is consistent with the X/P morphology and indicates that the X/P `bulge' and bar are aligned, since one has formed from, and is still the thick central part of, the other (see also \citealt{Martinez-Valpuesta&Gerhard2011}, \citealt{Romero-Gomez+2011} and \citealt{WeggGerhard&Portail2015}). There may be a small merger-built component to the Galactic bulge, with half light radius $R_e \approx 0.5$ kpc, assuming $h=2.54\pm0.16$ kpc (see \S \ref{sec:disc} in Appendix \ref{sec:mwdecomp}, where we model the Milky Way's radial light profile) and $R_e/h \approx 0.2$ (\citealt{CourteauDeJongBroelis1996}, \citealt{Graham&Worley2008}). However, we exclude the data in the inner 500 pc in \S \ref{sec:disc} and do not address the issue of a classical bulge in this paper, nor a nuclear bar, nor a nuclear disc (\citealt{Alard2001}, \citealt{Launhardt+2002}, \citealt{Nishiyama+2005}, \citealt{Gerhard&Martinez-Valpuesta2012}). Here we assume that strictly the X/P structure is aligned with the long bar and use it as a proxy for its orientation angle ($\alpha$ as above) as well as its extent.

From a sample of 88 galaxies with X--shaped bulges, \cite{Laurikainen&Salo2017} measured a mean $R_{{\it \Pi},{\rm obs}}/R_{\rm bar}$ ratio of $\approx$ 0.4, in good agreement with \cite{LuettickeDettmar&Pohlen2000}. The former authors, however, also found a subtle dichotomy in normalised (by bar length) sizes of X--shapes and those of barlenses, computing average ratios typically higher than $\gtrsim 0.5$ for barlenses. They concluded, based on the argument that X/P `bulges' and barlenses are the same structures viewed at different angles, that the intrinsic ratio is likely $\approx$ 0.5 for both (see their Fig. 8). More recently, \cite{Erwin&Debattista2017} place the mean of this ratio in the range $0.42 \leqslant R_{{\it \Pi},{\rm obs}}/R_{\rm bar}\leqslant 0.53$, where the lower and upper limits are determined by different definitions of bar length. With this in mind, based on the peak of the $B_6$ profile we estimate that the Milky Way bar has a radius of 4.2 $\pm 0.68$ kpc if the \Rpiint$/R_{\rm bar}$ ratio is 0.4, but may be as short as 3.2 kpc if \Rpiint$/R_{\rm bar} = 0.5$.


\subsection{X/P Diagnostics and Scaling Relations}\label{sec:results_scalingrelations}

The viewing angle of the Milky Way's X/P structure enables us to deproject the four radial $B_6$ profiles ($E$, $W$, $3.4\,\mu$m and $4.6\,\mu$m, Figure \ref{fig:b6_EW}), and thus compute the peanut's intrinsic metrics, such as length, height above the disc plane and integrated strength. The deprojected profiles (i.e., converted to a side-on view) are shown in Figure \ref{fig:b6_deproj}, along with an average profile (black curve) and its 1-$\sigma$ scatter (grey shaded region). Following CG16, we classify this as a `hump'--shaped profile which peaks at \Rpiint$ = 1.67$ kpc and declines to zero by $\approx$ 3 kpc. From the average, deprojected $B_6$ profile we computed the peanut's quantitative diagnostics, which are listed in Table \ref{tab:diagnostics}. 

Specifically, we report the maximum amplitude of the $6^{\rm th}$ order harmonic ($B_6$), labelled as ${\it \Pi}_{\rm max}$, the peanut intrinsic radius \Rpiint\ and height above the disc plane $z_{{\it \Pi},{\rm int}}$, the integrated strength of the peanut instability ($S_{\it \Pi}$), the full width at half-maximum of the $B_6$ signature ($W_{\it \Pi}$), as well as the qualitative shape of the $B_6$ profile, as used in CG16. Table \ref{tab:diagnostics} additionally reports the orientation angle ($\alpha$) of the (bar+X/P structure).

\begin{figure}\label{fig:b6_deproj}
\centering
	\includegraphics[width=1.02\columnwidth]{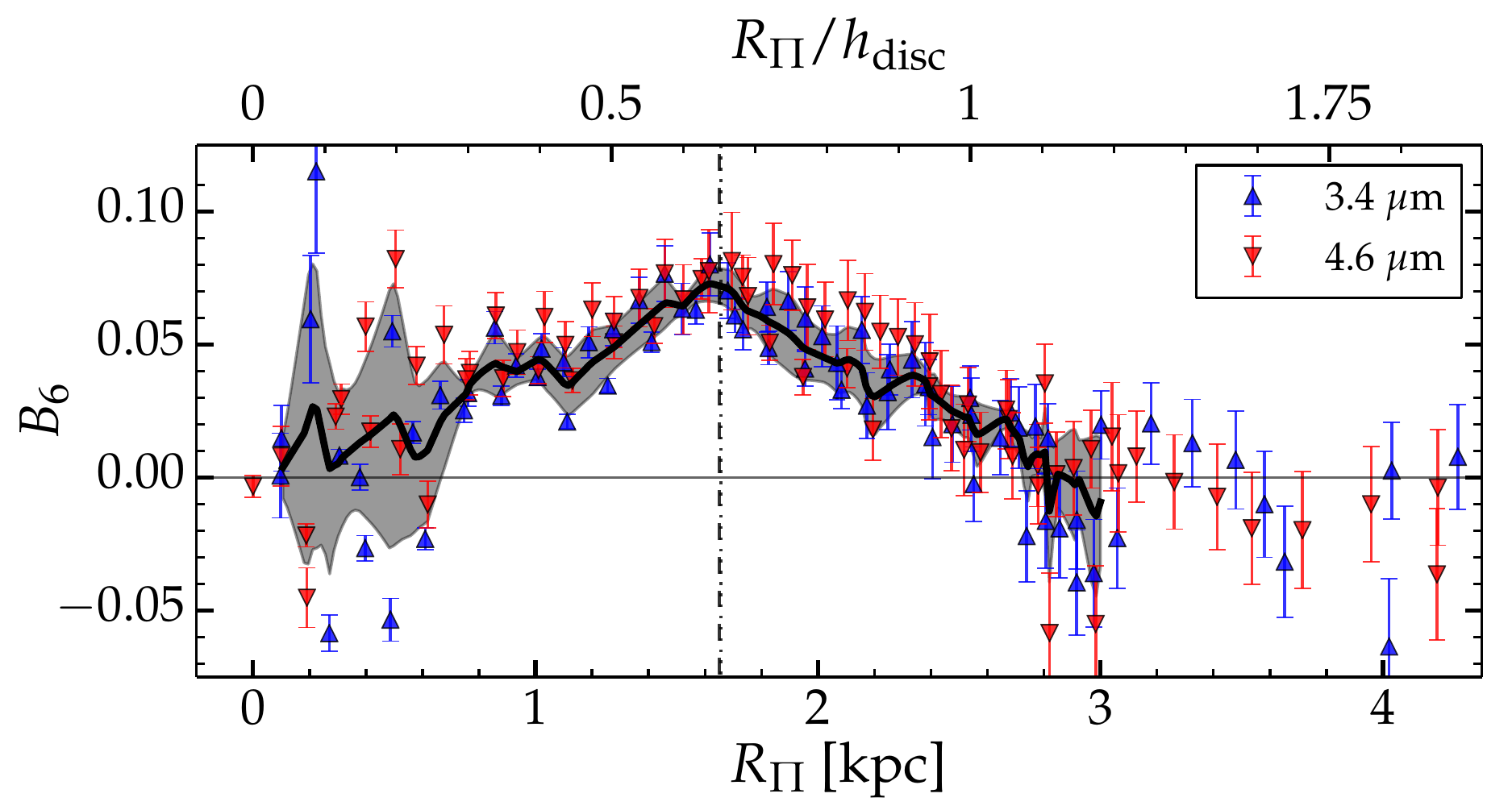}
	\caption{The radial $B_6$ profile of the Milky Way, as it would be viewed if the peanut were oriented side-on. The data points correspond to the extracted $B_6$ profiles in the $E$ and $W$ directions (Figure \ref{fig:b6_EW}), corrected for the bar's/peanut's viewing angle $\alpha$ (adjusted to a $90\degree$ orientation, rather than as observed at $37\degree$). The thick curve is the average over both directions and each wavelength, with the 1-$\sigma$ scatter shown through the shaded region. }
	\label{fig:b6_deproj}
\end{figure}  

\begin{figure}
\centering
\includegraphics[width=1.\columnwidth]{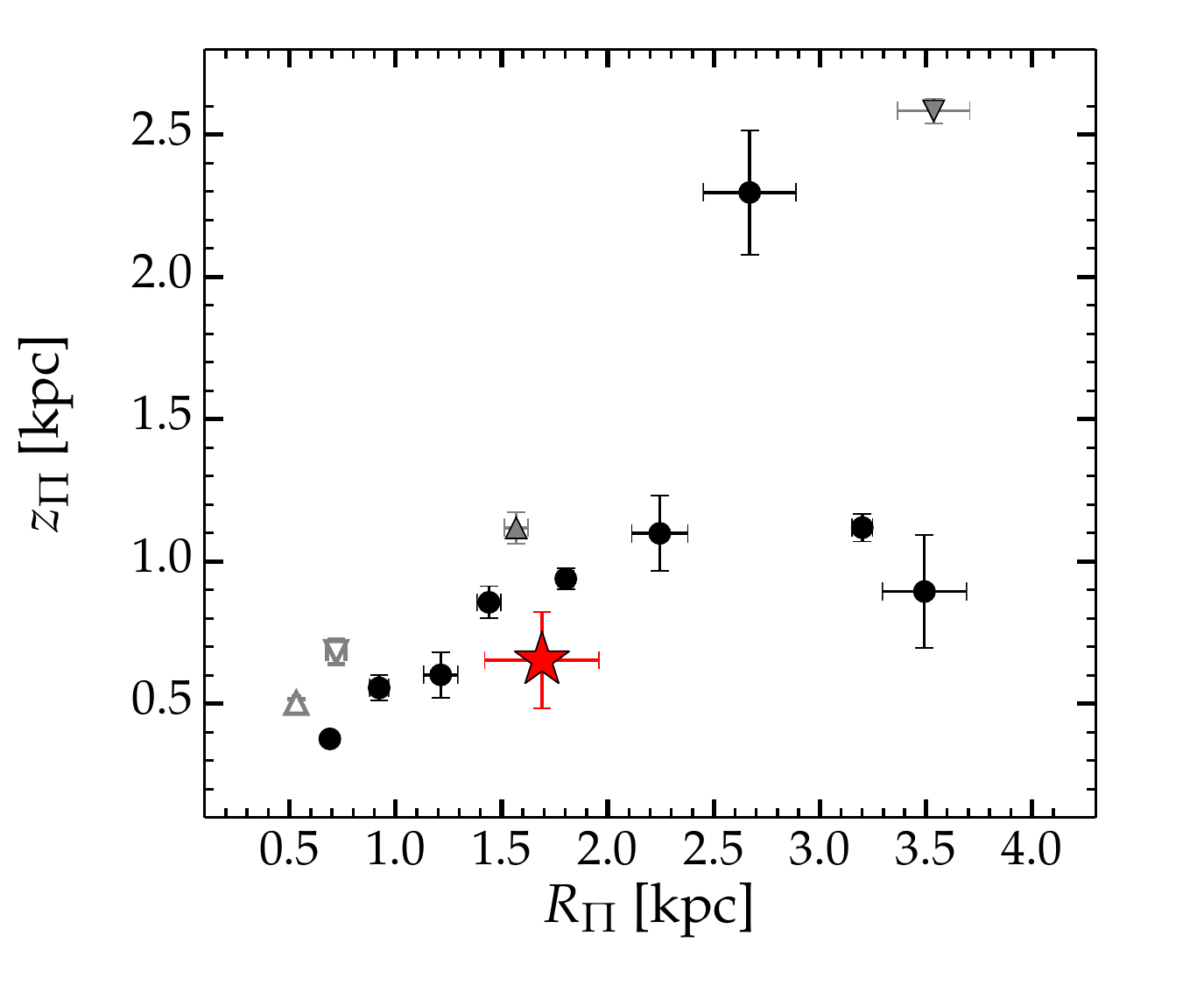}
\caption{Trend between X/P length within, and height above, the disc plane. Black and grey data from CG16, where $R_{\it \Pi}$ is projected and $z_{\it \Pi}$ is intrinsic. The red star is the Milky Way data point computed in this work, for which both $R_{\it \Pi}$ and $z_{\it \Pi}$ are intrinsic.}
\label{fig:z_vs_r}
\end{figure}

\begin{table*}
\centering
\begin{minipage}{60mm}
\caption{The Milky Way's X/P Diagnostics}
\centerline{
	\begin{tabular}{c c c c c c l}
	  \hline 
	  ${\it \Pi}_{\rm max}^{(a)}$ & $R_{{\it \Pi},{\rm int}}^{(b)}$ & $z_{{\it \Pi},{\rm int}}^{(c)}$ &  $S_{\it \Pi}^{(d)}$ & $W_{\it \Pi}^{(e)}$ & $\alpha^{(f)}$ & shape$^{(g)}$ \\
	   & [kpc,  units of $h$] &  [kpc, units of $h$] & [kpc, units of $h$] & [kpc, units of $h$] &[$\degree$] & \\
	  \hline \hline \\
	   0.073$\pm$0.007 & 1.67$\pm$0.27, 0.66$\pm$0.14 & 0.64$\pm$0.17, 0.25$\pm$0.07 & 5.67$\pm$2.00, 2.23$\pm$0.79 & 1.04$\pm$0.08, 0.41$\pm$0.04 & ${37} ^{+ 7}_{-10}$ & hump\\
	  \hline 
	\end{tabular}\\	
}
\label{tab:diagnostics}
\end{minipage}
\begin{minipage}{169mm}
$(a)$-- maximum amplitude of $B_6$ harmonic; $(b)$-- intrinsic radius of X/P structure; $(c)$-- intrinsic vertical height of X/P structure; $(d)$-- integrated strength of the $B_6$ profile; $(e)$-- full width at half-maximum of the $B_6$ profile; $(f)$-- peanut angle with Sun-(Galactic Centre) line-of-sight; $(g)$-- qualitative shape of the $B_6$ profile (as defined in CG16).
\end{minipage}
\end{table*}

CG16 have shown that the X/P parameter space is not randomly populated but rather the X/P metrics give rise to several scaling relations. One such correlation involves the peanut radius, $R_{\it \Pi}$,  and its vertical height above the disc, $z_{\it \Pi}$. This is shown in Figure \ref{fig:z_vs_r}, where the black and grey data points correspond to the twelve galaxies in the CG16 sample\footnote{The four grey data points correspond to two galaxies with nested X/P structures: hollow symbols for the inner and filled symbols for the outer.}, and the red star corresponds to the Milky Way value as obtained here. This trend is relevant for constraining the age of X/P structures, in light of their `radial drift' (see e.g., \citealt{Quillen+2014}). As the peanut is believed to arise at the inner Lindblad resonance point, the bar's slowing down causes the resonance point to drift outward, elongating the peanut.

The Milky Way is consistent with the general trend in Figure \ref{fig:z_vs_r}, though appears to be marginally shifted towards a slightly higher $R_{\it \Pi}$ value (or lower $z_{\it \Pi}$). However, in their analysis, CG16 were limited by the unknown viewing angles of the galactic bars in their galaxy sample, and hence their measured X/P radii were in fact projected quantities, i.e., their data are $R_{\it \Pi} \equiv R_{{\it \Pi},{\rm obs}} \leqslant R_{{\it \Pi},{\rm int}}$. For the Milky Way, our determination of the bar's viewing angle relieves this limitation and so our X/P radius is intrinsic, i.e. $R_{\it \Pi} \equiv R_{{\it \Pi},{\rm int}}$. Note that CG16 obtained intrinsic $z_{\it \Pi}$ values by using the inclinations of the galaxy discs to correct for projection effects in the vertical direction. Our  $z_{\it \Pi}$ value is also intrinsic, since we are viewing the Galaxy's disc almost perfectly edge-on (the disc's inclination is $i \lesssim 0\degree\!.2$). 

\begin{figure*}
\includegraphics[width=.9\columnwidth]{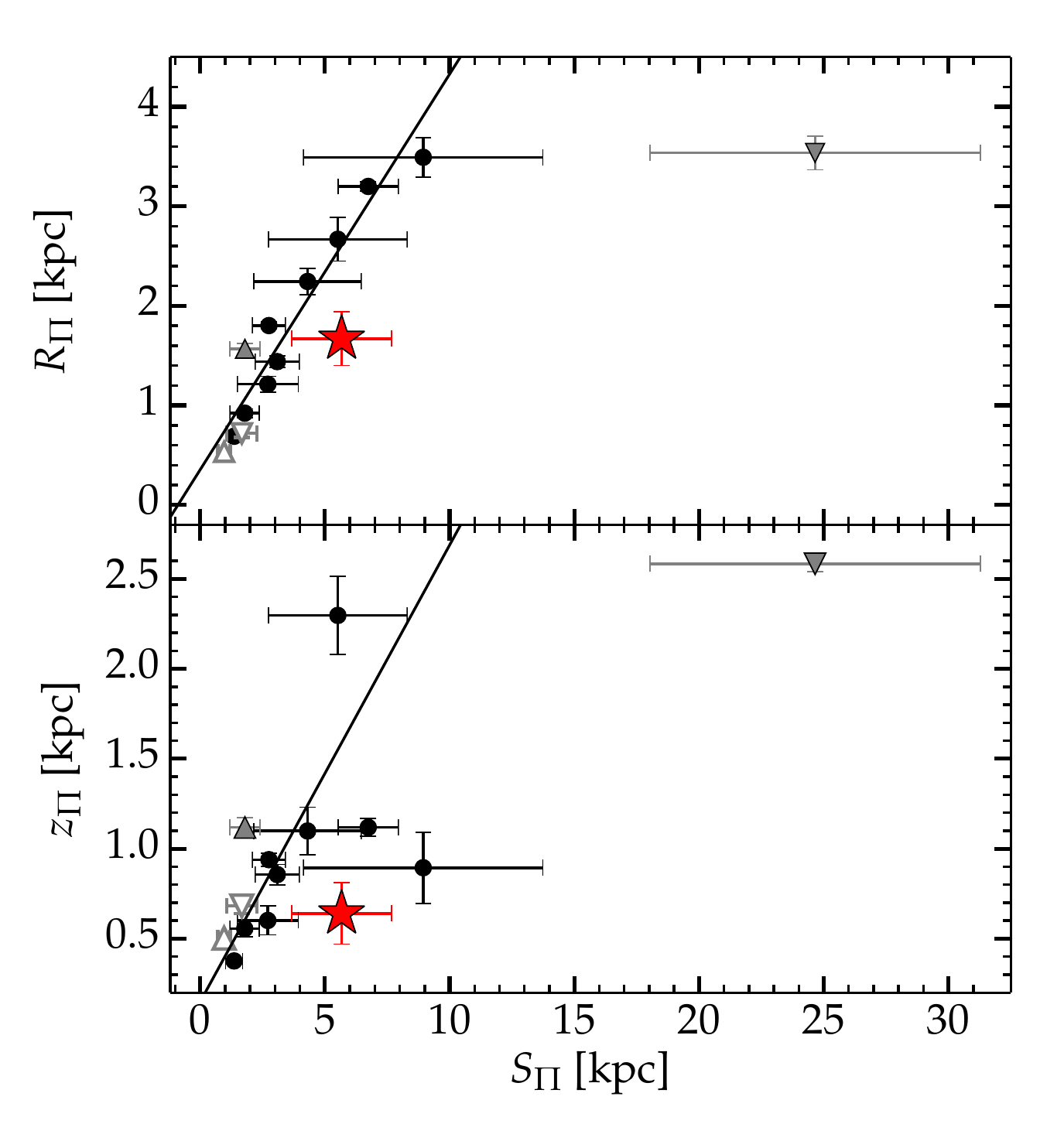} \includegraphics[width=.9\columnwidth]{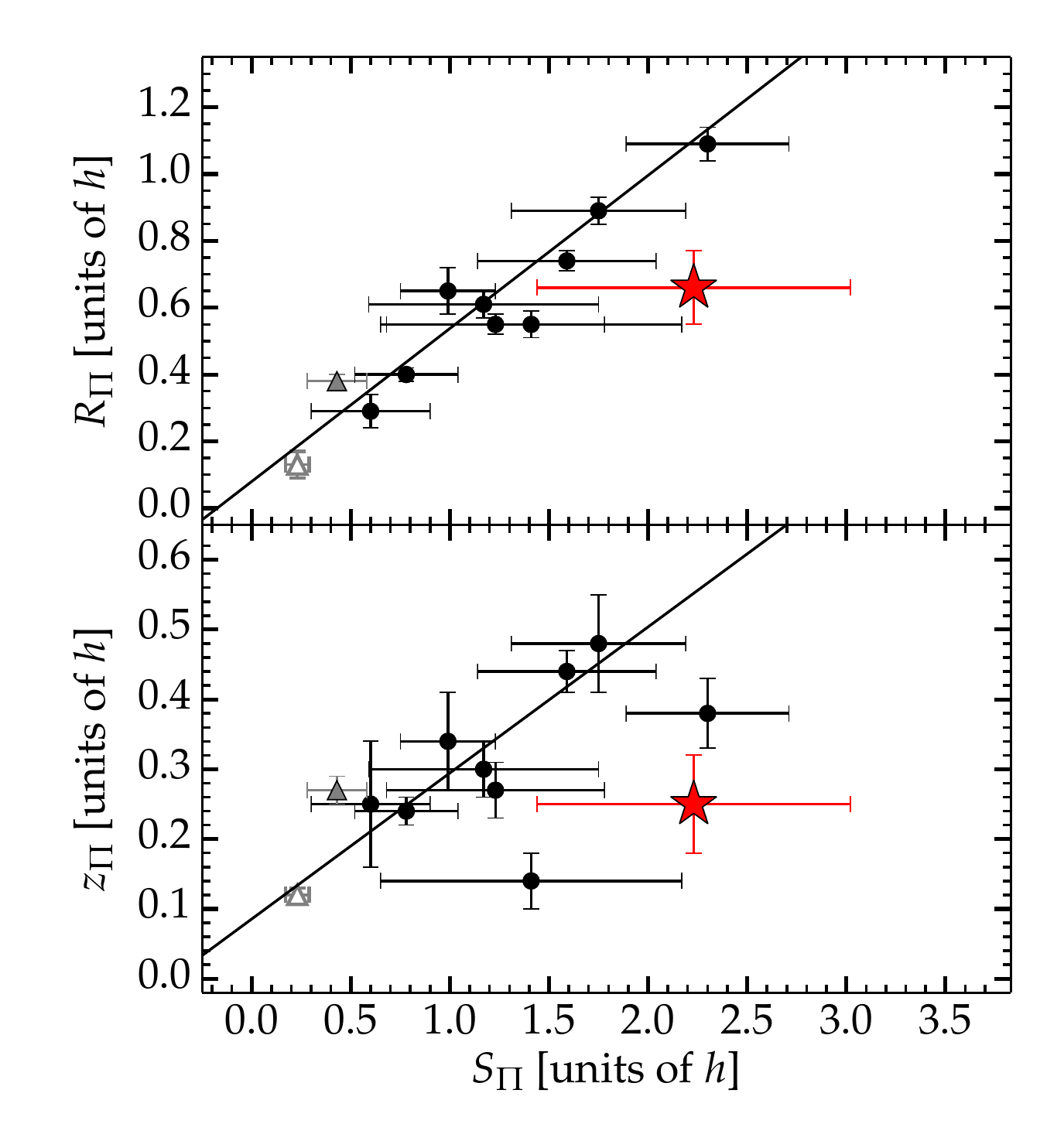}
\caption{CG16 scaling relations showing X/P radius (top) and height (bottom) as a function of integrated strength. The colour scheme is analogous to Figure \ref{fig:z_vs_r} and the lines represent linear fits from CG16. The correlations are shown in kpc (left) and in units of disc scale length $h$ (right). The outer peanut of NGC 128 is an outlier from the trends (outside the plotting area in the right-hand panels), possibly having its X/P strength enhanced through interactions with its satellite.}
\label{fig:zR_vs_S}
\end{figure*}

\begin{figure*}
	\centering
	\includegraphics[width=0.49\textwidth]{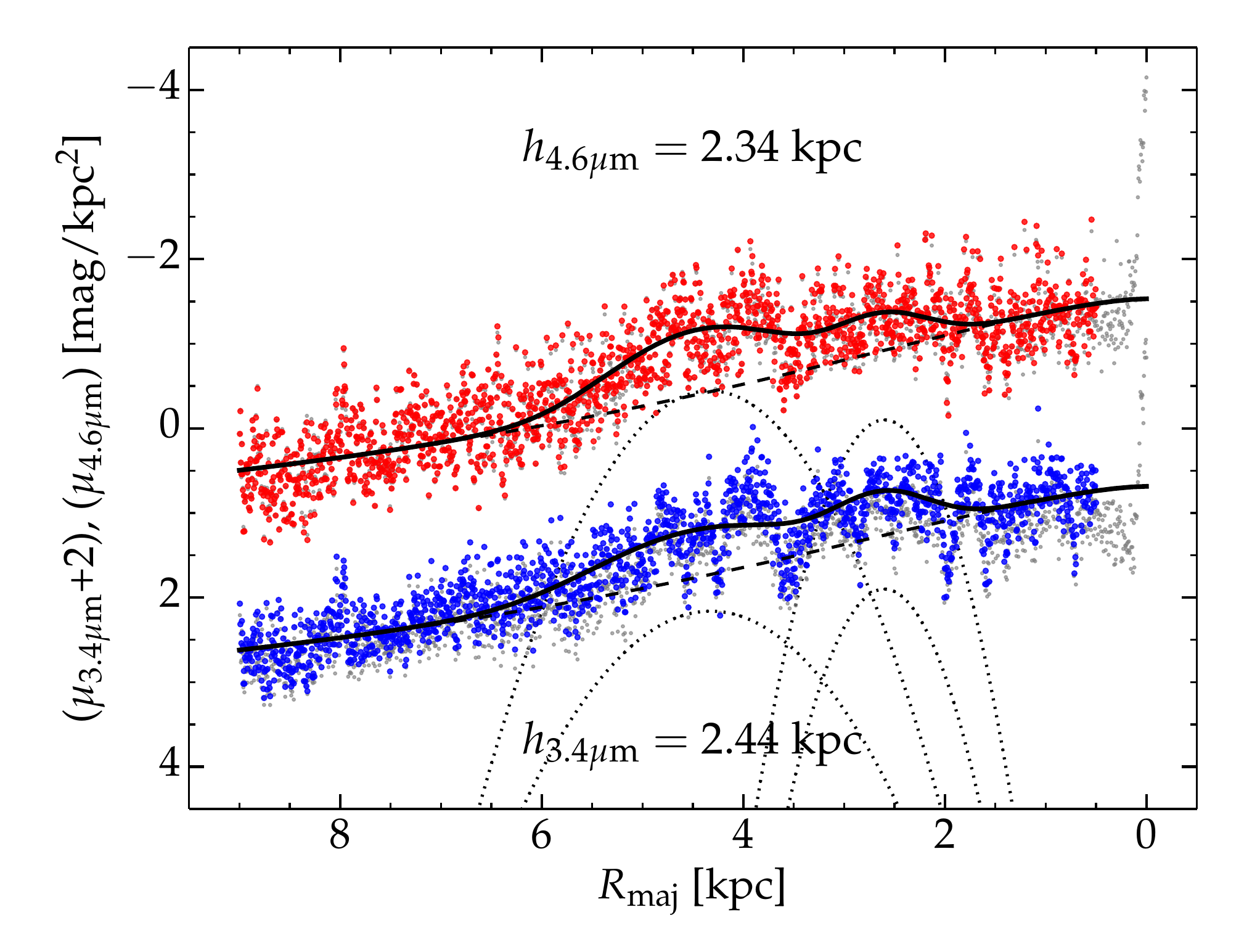}
	\includegraphics[width=0.49\textwidth]{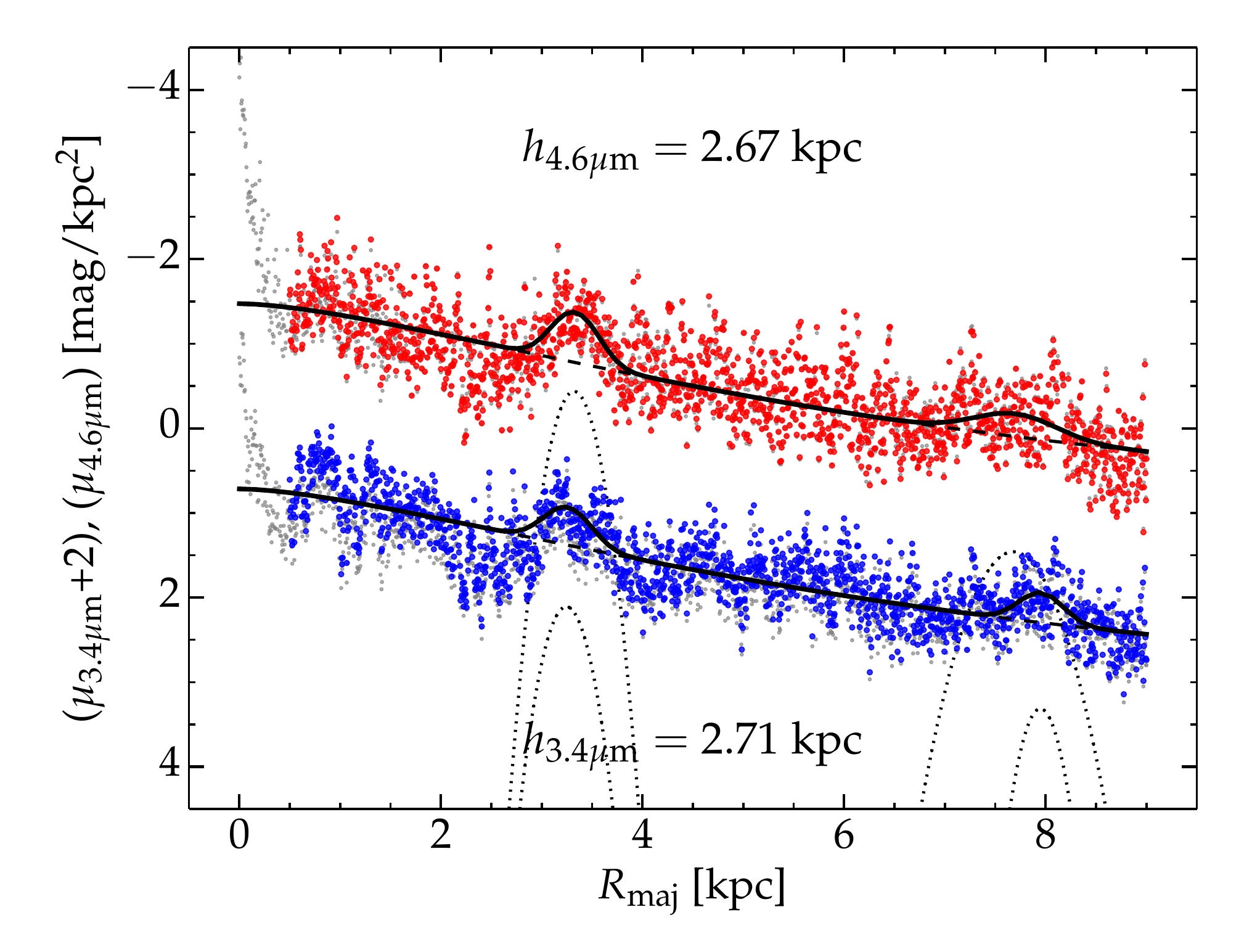}\\
	\caption{1D cuts in the plane of the disc to the $East$ of the Galactic Centre (left-hand side) and to the $West$ (right-hand side). Grey symbols represent raw cuts from processed images (as in Figure \ref{fig:wise}) while blue and red data are corrected for dust extinction and glow and correspond to the $3.4\,\mu$m and $4.6\,\mu$m data, respectively. Black curves represent the best-fitting model (exponential+2 Gaussians), corrected for our vantage point within the disc and assuming Sun's Galactocentric distance of 8.2 kpc. Insets indicate the best-fit disc scale length $h$ for each panel. The $3.4\,\mu$m profiles are offset by 2 magnitudes, for display clarity, and the inner 500 pc were excluded from the fits, since the light in that radial range is dominated by a small scale but bright component.} 
	\label{fig:discfit_best}
\end{figure*}

Another set of correlations occur between the X/P size (length and height) and its integrated strength $S_{\it \Pi}$ (Equation \ref{equ:sharpness}). These are shown in Figure \ref{fig:zR_vs_S}, where, as before, the black and grey data corresponds to the CG16 sample. The line is their linear fit to the data and the red star corresponds to the Milky Way. Interestingly, these trends also hold when plotted in units of the disc's scale length (rather than in kpc), indicating that peanuts `know' about their host disc. CG16 proposed to normalise, where applicable, the metrics of the peanut structures by $h$, since this provides quantities that are independent of the type or size of individual galaxies, or the uncertainties in their distance estimates. This also facilitates comparisons with numerical simulations. We determined the scale length of the Milky Way by performing a photometric decomposition of the major axis surface brightness profile, separately in the $E$ and $W$ directions, and taking into account the Sun's placement within the disc as well as the Galaxy's spiral structure. The full analysis is presented in Appendix \ref{sec:mwdecomp}. Our preferred models, shown in Figure \ref{fig:discfit_best}, resulted in an average value over both bands and both directions, of $h=2.54\pm0.16$ kpc, in good agreement with the literature (\citealt{Licquia&Newman2016}, \citealt{Bland-Hawthorn&Gerhard2016}).

Figure \ref{fig:zR_vs_S} shows how the Milky Way fits in with the ($z_{\it \Pi} - S_{\it \Pi}$) and ($R_{\it \Pi} - S_{\it \Pi}$) scaling relations. The Galaxy is consistent (within 2$\sigma$) with the trend seen in the CG16 sample, albeit with an X/P strength $S_{\it \Pi}$ that is somewhat on the high side. The X/P strength, however, is also sensitive to the bar viewing angle $\alpha$, since $S_{\it \Pi}$ is an integral of the $B_6$ curve and $\alpha$ controls the deprojection (`stretching'), of the $B_6$ profile when adjusting to a side-on orientation of the peanut (Figure \ref{fig:b6_deproj}). As $\alpha$ was unknown for the CG16 galaxies, the scaling relations presented are between projected, and thus potentially underestimated in-plane quantities.

\begin{figure*}
\includegraphics[width=.75\textwidth]{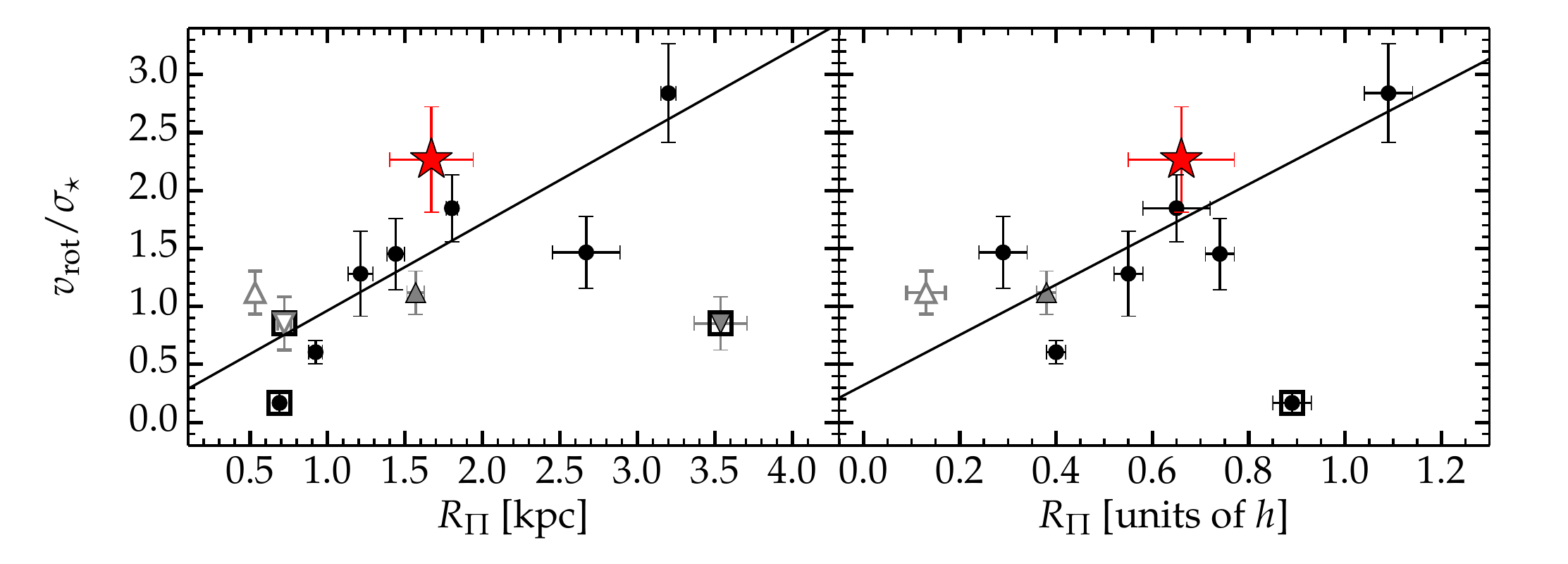}\\
\includegraphics[width=.75\textwidth]{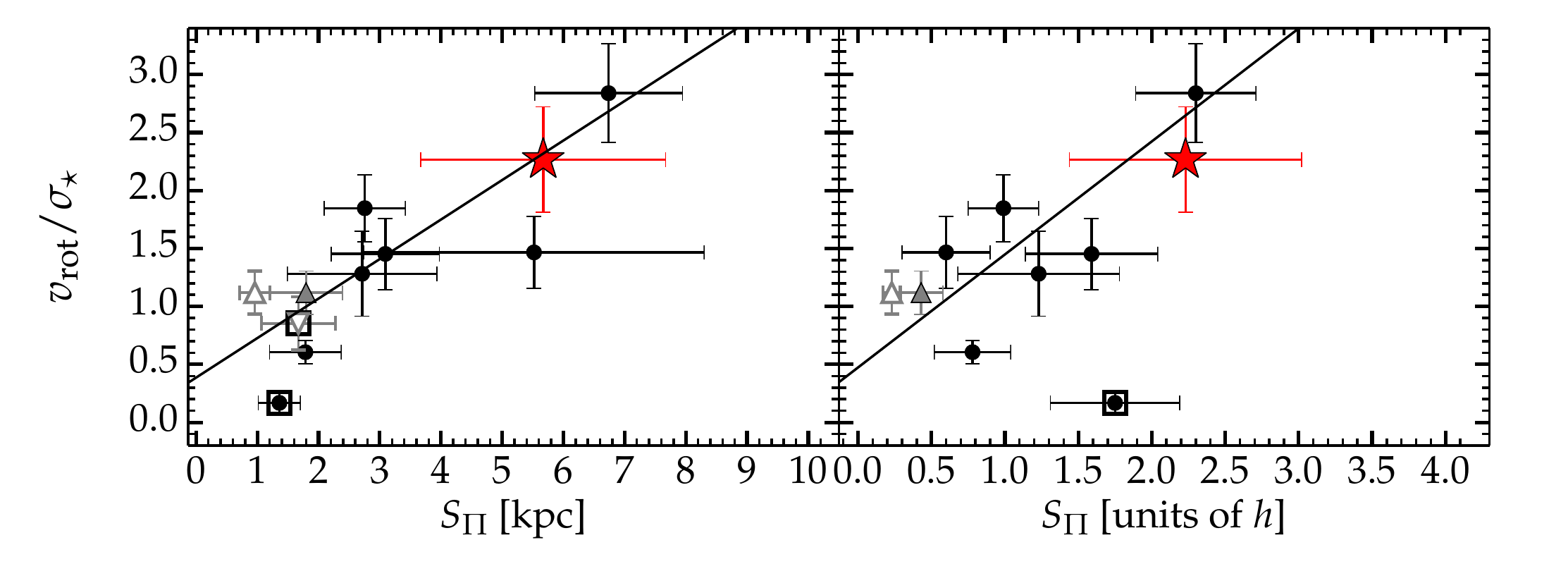}
\caption{CG16 scaling relations between galaxy $v_{\rm rot}/\sigma$ and the peanut properties: radius (top) and strength (bottom). The colour scheme is analogous to Figure \ref{fig:zR_vs_S}, and data points framed in squares were excluded from the fit in CG16 (see \S \ref{sec:results_scalingrelations}). The correlations hold when the X/P parameters are both in kpc (left) and in units of disc scale length $h$ (right).}
\label{fig:vSig_RS}
\end{figure*}

Finally, X/P structures are also known to correlate with their host galaxy's kinematics (\citealt{Bureau&Freeman1999}, \citealt{Debattista+2005}, \citealt{Ianuzzi&Athanassoula2015}, \citealt{AthanassoulaRodinov&Prantzos2017}). CG16 have shown a (weak) trend between galaxy $v_{\rm rot}/\sigma$ (rotation velocity/velocity dispersion) ratio and the length and strength of the peanut structures, such that larger and stronger peanuts occur in more rotation-dominated systems. These correlations are shown in Figure \ref{fig:vSig_RS}, where the colour scheme is analogous to Figures \ref{fig:z_vs_r} and \ref{fig:zR_vs_S}. The data points framed in open squares have unreliable $v_{\rm rot}/\sigma$ ratios (see CG16 for details). As in Figure \ref{fig:zR_vs_S}, these correlations also hold when the X/P parameters are normalised by the disc scale length $h$, once again indicating that the disc in which peanuts are embedded is important. For the Milky Way we adopted a $v_{\rm rot}/\sigma$ ratio of $2.27\pm0.44$ based on a disc rotation velocity of 238$\pm$15 km\;s$^{-1}$ (\citealt{Bland-Hawthorn&Gerhard2016}; see also \citealt{Schoenrich2012}, \citealt{Reid+2014}, \citealt{Reid&Dame2016}) and a central velocity dispersion of $105\pm20$ km\:s$^{-1}$ (\citealt{Merritt&Ferrarese2001}, \citealt{Gueltekin+2009}). 

\section{Discussion}\label{sec:discussion}

\subsection{The Milky Way's X/P Parameters in Context}\label{sec:discus_XP}

The spatial parameters (length, height above the disc) of the Milky Way's X/P structure measured in this paper agree well with those of other nearby galaxies, making our Galaxy typical in this respect. The integrated strength of the X/P structure appears, however, to be moderately larger than the general trend, which may be due to projection effects, as explained in \S \ref{sec:results_scalingrelations}. Specifically, the peanut strength, $S_{\it \Pi}$, is sensitive to the orientation angle ($\alpha$) at which the bar, and X/P structure, are viewed. In a more end-on orientation, the observed (in projection) $B_6$ profile is more `contracted' compared to a side-on view, and as the integral over this profile, $S_{\it \Pi}$ has a maximal value in side-on orientation and decreases with decreasing $\alpha$. While in this work our knowledge of $\alpha$ allowed us to deproject the Milky Way's $B_6$ profile to side-on orientation, the galaxies in CG16 had unknown bar/peanut viewing angles, and hence possibly underestimated $S_{\it \Pi}$ values. Note that an unknown $\alpha$ would also imply potentially underestimated $R_{\it \Pi}$ values, but would not bias the peanut height ($z_{\it \Pi}$) measurements, which in CG16 are intrinsic values. Therefore, projection effects may only explain the moderate offset of the Milky Way in the $z_{\it \Pi} - S_{\it \Pi}$ trends (bottom panels in Figure \ref{fig:zR_vs_S}).

An alternative, and intriguing, explanation for this is that the Milky Way may have had its X/P strength enhanced through tidal interactions with its infalling satellites, such as the $Small$ and {\it Large Magellanic Cloud}, or the disrupted $Sagittarius$ dwarf (\citealt{Jiang&Binney2000}). Attempting to explain how boxy/peanut/X--shaped structures form, \cite{Binney&Petrou1985} and \cite{Rowley1988} argued that interactions with small satellite galaxies  (disruption and accretion of material) can give rise to orbit families that lead to rectangular, boxy isophotes and cylindrical rotation in their larger companions. While this scenario was ruled unlikely to be the primary formation mechanism of X/P structures (see \citealt{Bureau&Freeman1999}, their Sec. 2.1), satellite interactions may still serve to enhance the strength of the peanut. For example, NGC~128, one of the most prominent X/P galaxies, clearly shows material exchange with its smaller companion NGC~127, as shown in Fig. 3 in CG16. By contrast, the rest of the CG16 sample of X/P galaxies did not show any clear evidence of satellites. As such, the datum corresponding to NGC~128\footnote{More precisely, to the $outer$ peanut of NGC~128. The inner peanut (empty grey downward triangle in Figure \ref{fig:zR_vs_S}) appears to fit the trend quite well.}, plotted as the filled grey downward triangle in Figure \ref{fig:zR_vs_S}, is a significant outlier of the trend. Note that accretion of the intergalactic medium (\citealt{Lopez-Corredoira+2002}) may also play a role in this respect.

Interestingly, the Milky Way's isophotes in the X/P region show an apparent, though weak, $North-South$ asymmetry, such that the  northern two `arms' of the X shape appear to have a wider opening angle than the southern two arms, in both filters. This is reminiscent of bars in the buckling phase seen in simulations (e.g., \citealt{Martinez-ValpuestaShlosman&Heller2006}) as well as observations (e.g., \citealt{Erwin&Debattista2016}), which is the primary instability mechanism that leads to X/P structures. We may be observing the remaining signature of the Milky Way's past bar buckling event. The asymmetry, however, is only apparent on the eastern (closer) limb of the peanut structure (Figure \ref{fig:reflections}, panels $a$ and $c$), which warrants a more in-depth study of differences between positive and negative latitudes. This is, however, beyond the scope of this paper.

\subsection{The Long Bar Parameters: Implications}\label{sec:discus_barLit}

\subsubsection{Comparison with Literature}\label{sec:discus_Bar_comparison}

In Figure \ref{fig:alpha_rbar} we compare our bar parameters (orientation angle and radius) with other results from the literature. Our preferred parameters of $\alpha = {37\degree}^{+ 7\degree}_{-10\degree}$ and $R_{\rm bar} = 4.16 \pm 0.68$ kpc agree well with \cite{Zasowski2012}, who measured $\alpha = 38\degree\pm6\degree$ from \GLIMPSE\ (\citealt{Benjamin+2005}, \citealt{Churchwell+2009}) data, and the recent study of \cite{Monari+2017}, who show evidence for a relatively short and fast bar with a co-rotation radius of $\sim 4$ kpc. We plot our preferred parameters, which assume a \Rpiint$/R_{\rm bar}$ ratio of 0.4, in Figure \ref{fig:alpha_rbar} as the red star symbol. Additionally, our lower estimate for the bar length, which assumes \Rpiint$/R_{\rm bar} = 0.5$, is shown by the black star symbol. The literature results were taken from \cite{Picaud2004} (P04; $\alpha = 45\degree\pm9\degree$, $R_{\rm bar} = 3.9\pm0.4$ kpc), \cite{Benjamin+2005} (B05; $\alpha = 44\degree\pm10\degree$, $R_{\rm bar} = 4.4\pm0.5$ kpc), from the combined works of the group \cite{Hammersley+2000}, \cite{Lopez_Corredoira+01,Lopez-Corredoira+2006} and \cite{Cabrera-Lavers+2007, Cabrera-Lavers+08} (HLC; $\alpha = 43\degree\pm\sim\!2\degree$, $R_{\rm bar} = 3.9-4.5$ kpc), from \cite{Francis&Anderson2012} (FA12; $\alpha = 30\degree\pm10\degree$, $R_{\rm bar} = 4.2\pm0.1$ kpc) and from \cite{WeggGerhard&Portail2015} (WGP15; $\alpha = 28\degree-33\degree$, $R_{\rm bar} = 4.6\pm0.3-5.0\pm0.2$ kpc). Our preferred data point, without considering the error bars for the moment, is consistent (within the errors) with P04, B05 and FA12, but appears to show tension with WGP15 and HLC, i.e., lying roughly between their respective ranges but outside their uncertainty intervals, which are comparatively smaller than the other studies and, notably, exclude each other. The latter two groups advocate competing interpretations of the Milky Way's central components. HLC posit the existence of a long thin bar and a shorter, thicker, triaxial bulge, the two misaligned with each other. WGP15 on the other hand advocate the notion that the long bar has a smaller orientation angle, and is thus aligned with the X/P structure, and that in fact the latter is essentially the central, vertically thickened part of the former.

\begin{figure}
\centering
\includegraphics[width=1.\columnwidth]{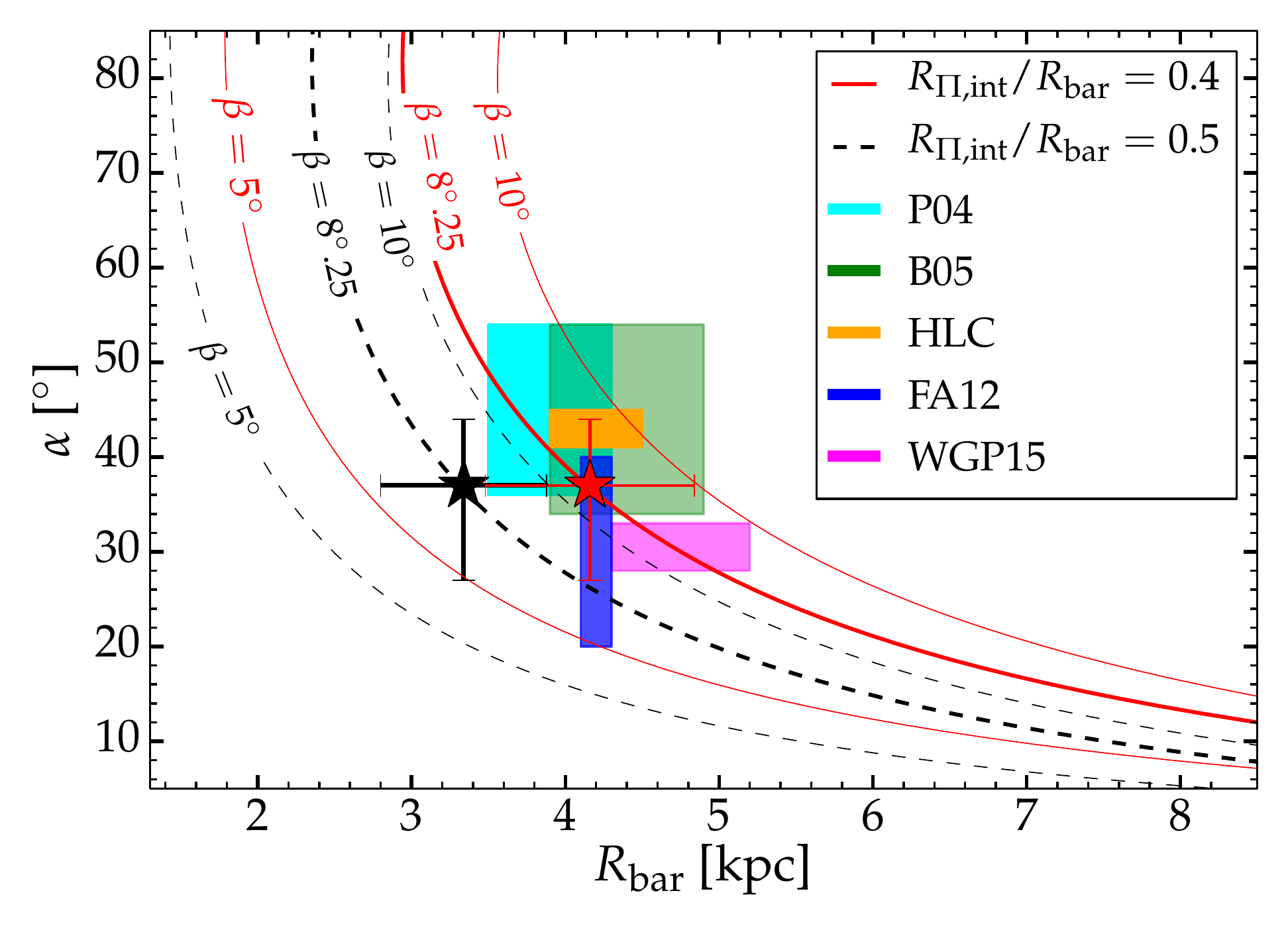}
\caption{Bar radius vs. orientation angle $\alpha$. Curves illustrate the coupling of the two parameters given our $\beta$ (angular size of the peanut eastward of the Galactic Centre) measurement (thick) and taking reasonable upper and lower limits of it (thin). Red solid and black dashed curves assume different \Rpiint$/R_{\rm bar}$ ratios (see legend). Boxes indicate literature results and their uncertainties, while the stars are the results of this work, assuming \Rpiint$/R_{\rm bar}$=0.4 (red) and 0.5 (black).}
\label{fig:alpha_rbar}
\end{figure}

Due to our substantial uncertainty intervals, our result does not rule out either of the above two scenarios. But were we to relax some of our assumptions or measurements, and explore the systematics and sources of uncertainty in our analysis, could we arrive at a better agreement with either of the two pictures? We explore this in the following sub-sections, by again looking at the ($\alpha - R_{\rm bar}$) parameter space.

\subsubsection{Limitations and Systematics}\label{sec:discus_Bar_limitations}

Although our methodology for detecting X/P structures is both sensitive and accurate for external galaxies (capable of detecting `nested' X/P structures, as shown in CG16), our vantage point of the Milky Way may introduce uncertainties in this analysis. Specifically, we are observing the X/P structure through intervening disc light, which may `wash out' the faint extremities of the peanut, both in--plane and in the vertical direction. A more accurate approach would involve the use of data that is not affected by disc light, e.g., (2D) maps of the distribution of RCG stars, which are commonly used as tracers of Galactic structure. In addition, our analysis only considered the radial (length) and vertical (height) directions of what is in fact a three-dimensional structure. Additional uncertainties in the true `ends' of the peanut may arise from its in-plane `thickness', and how this projects onto the plane of the sky (e.g., Fig. 6 in \citealt{Lopez-Corredoira+2006}; see also \citealt{Buta&Crocker1991}, \citealt{Buta1995}, \citealt{Laurikainen+2011} and \citealt{Salo&Laurikainen2017} for interesting examples of peanuts viewed face-on). To avoid most of the aforementioned issues, and keep consistency with CG16, we have used the peak in the $B_6$ profile, rather than the point where it declines to zero, as the indicator of the peanut's characteristic scale. At this point the peanut is most prominent, and hence using it additionally ensures consistency with other studies that have measured X/P structures, which relied on identification techniques (e.g., visual inspection, unsharp masking) that are sensitive to the point where the feature is most prominent. 

Of particular interest for this paper are studies which report the typical value of $R_{\it \Pi}/R_{\rm bar}$, since we have relied on this ratio to obtain the bar length. Recent studies place its mean value, in nearby X/P galaxies, between $\sim 0.4 - 0.5$ (\citealt{Laurikainen&Salo2017}, \citealt{Erwin&Debattista2017}), but all find scatter in it. Prima facie, our analysis shows that a value closer to 0.4 for the Milky Way is more consistent with the bar parameters in the literature, while a value of 0.5 appears to underestimate the bar length (Figure \ref{fig:alpha_rbar}). However, in the following sub-section we investigate how the reliability of our measured X/P size, and how the applicability of the \Rpiint$/R_{\rm bar}$ ratio to our measurements of the Milky Way, affects our results.

\subsubsection{Exploring the ($\alpha - R_{\rm bar}$) Coupling}

Considering that we observe the (bar+X/P structure) in projection, it is obvious that our derived intrinsic X/P radius \Rpiint\ (and, by extension, $R_{\rm bar}$) and viewing angle, are correlated quantities: a given $projected$ size (i.e., the measurement/observation) can correspond to a large intrinsic size if the viewing angle $\alpha$ is small, or to a smaller intrinsic size if the angle is larger (see Figure \ref{fig:geometry}, which applies to both the peanut and the bar, and any elongated structure viewed at an angle). This ($\alpha$ -- intrinsic size) coupling, is shown in Figure \ref{fig:alpha_rbar} through the red and black curves, for which the observed quantity (projected size) is $\beta$, i.e., the peanut's angular size in the eastern direction (see Figure \ref{fig:geometry}). If we were to assume that our measured value of $\beta = 8\degree\!.25$ is the only information we have\footnote{We chose $\beta$ because it corresponds to the nearer limb of the peanut, which in principle should be easier to measure. However, we repeated the exercise with $\gamma$ -- the projected angular size on the $West$ (far) side -- and obtained similar results.}, then the data point must lie on the thick red curve, if \Rpiint$/R_{\rm bar} = 0.4$ (our preferred scenario), or on the thick dashed curve if \Rpiint$/R_{\rm bar} = 0.5$. If we assume that the true value of $\alpha$ is smaller than $37\degree$ (i.e., if we assume that our measurement of $\gamma$ was biased, since $\beta$ and $\gamma$ together constrain $\alpha$), and is more in the region of $\sim 30\degree$, then travelling down the thick red curve brings us in good agreement with WGP15. On the other hand, a higher value of $\alpha$  ($\sim 43\degree$) improves the agreement with HLC. If we further assume that our measurement of $\beta$ was biased as well, and the true end of the peanut occurs beyond $8\degree\!.25$ (we show 10$\degree$ in Figure \ref{fig:alpha_rbar}, a typical upper limit for the bulge--bar transition), then the opposite occurs. A lower value of $\beta$ increases the discrepancy with all the literature numbers. All of this however is for a fixed \Rpiint$/R_{\rm bar}$, a ratio necessary to map the X/P size ($\beta$) onto a bar size. Varying this ratio translates the three red curves in the $x$--direction, as illustrated through the black dashed curves, which are equivalent to the red curves but for a higher \Rpiint$/R_{\rm bar}$ value of 0.5.  

Most studies report on a bar length $\gtrsim 4-4.5$ kpc, which, in conjunction with our work, suggest that for the Milky Way, \Rpiint$/R_{\rm bar}$ is close to $\approx 0.4$. However, the long bar may not be as long after all. In a recent paper, \cite{Monari+2017} argue, based on $Gaia$ DR1 (\citealt{Gaia2016}) and LAMOST (\citealt{Liu+2014}) data, that the position of the $Hercules$ stream in velocity space favours a shorter bar, with a co-rotation radius of $\sim\!4$ kpc (at odds with \citealt{Portail+2017}, who report a longer, $\sim\!6$ kpc radius of co-rotation). A shorter bar would also be more consistent with bar-to-disc sizes in other disc galaxies, as the Milky Way is usually invoked to be a typical barred spiral. \cite{Erwin2005} found bar sizes to range between 1--10 kpc (with a mean of 3.3 kpc) or 0.5--2.5\,$h$ for early-type disc galaxies (S0--Sab). Later-type disc galaxies, such as the Milky Way, which is believed to be Sb or Sbc, by most sources (\citealt{Hodge1983}, \citealt{Kennicutt2001}), have comparatively shorter bars, ranging from 0.5--3.5 kpc, or 0.2$h$--1.5$h$. Assuming our measured value of $h=2.54$ kpc for the disc's exponential scale length, this maps the WGP15 range (4.6--5) kpc into (1.8--2)\,$h$, the HLC range (3.7--4.5) kpc into (1.5--1.8)\,$h$ and our estimated range of (3.3--4.2) kpc into (1.3--1.7)\,$h$. Naturally, these numbers carry quite large uncertainties not only due to intrinsic scatter but also due to different definitions of `bar length' (see \citealt{Athanassoula&Misiriotis2002}, their Sec. 8). 

As previously mentioned, \cite{Laurikainen&Salo2017} report a mean $R_{{\it \Pi},{\rm obs}}/R_{\rm bar}$ ratio of $\sim\!$ 0.4 for X/P structures while for barlenses their measurements exceed $\sim\!$ 0.5. From the argument that X/P structures and barlenses are the same structures viewed at different inclinations (edge-on vs face-on) and by analysing simulated X/P galaxies at different projection angles, they conclude that the mean intrinsic ratio is $\approx$ 0.5 for both features (with some scatter). While most literature measurements of the length of the long bar, coupled with our \Rpiint, favour an \Rpiint$/R_{{\it \Pi},{\rm bar}}$ ratio of $\approx$ 0.4 for the Milky Way, a value closer to 0.5 would imply a shorter bar, as seen in Figure \ref{fig:alpha_rbar} (black star symbol). A shorter bar is not necessarily in contradiction with the findings of many authors. As suggested by \cite{Monari+2017}, a flat stellar distribution extending further than 4 kpc could simply correspond to loosely wound spiral arms that originate from the bar's ends. In light of the above arguments, we choose to keep our shorter estimate of $R_{\rm bar} = 3.24\pm0.54$ kpc as a plausible value.

The scenario in which the Milky Way's `bulge' is the inner, thickened, X/peanut-shaped region of its long bar, which has arisen through the buckling of the former (\citealt{Martinez-Valpuesta&Gerhard2011}, \citealt{Romero-Gomez+2011}, WGP15), is a natural interpretation of our Galaxy's central components. This scenario is supported by numerical simulations as well as observational evidence that most of the stars in the bulge originate from the disc (\citealt{Shen+2010}; \citealt{Ness+2013,Ness+2014}; \citealt{DiMatteo+2014,DiMatteo2016}), implying that it formed predominantly from the buckling and secular evolution of the disc and bar. In support of this picture, WGP15 have argued that the angle of the long bar is smaller than previously thought, and is consistent with that of the elongated `bulge'. While we agree with WGP15 that the two structures are likely aligned, we propose, and show evidence, that it is not the long bar which has a lower angle ($\sim\!30\degree$) than most literature measurements but that the X/P `bulge' instead has a larger angle ($\sim\!37\degree$) than previously thought. If WGP15 increase their $\alpha$ value to our value of $\sim37\degree$ (i.e., move up the red curve in Figure \ref{fig:alpha_rbar}), then their result would agree with our work and produce a bar radius shorter than 5 kpc.

\subsection{The End of the Bar}

An accurate accounting of the long bar is crucial if we are
to understand the inner dynamics of the galaxy and, in particular,
the disc-bar-bulge transition in this region. This has been a 
long-standing problem in the widely used Besan\c{c}on (\citealt{Robin+2003})
and Galaxia (\citealt{Sharma+2011}) models of the Galaxy, for example.
At present, these inner structures are inserted artificially and
do not conform to a dynamically self-consistent framework.

\citealt{WeggGerhard&Portail2015} have revealed that there are two scale height components
extending into the long bar region: the `thin' component and the `superthin' component. The `thin` bar has a scale height of 180 pc, with a declining density with radius, and appears to be the 
barred counterpart of the old inner disc. The `superthin' component has
a remarkably small scale height of 45 pc, and the density appears to
{\it increase} outwards. They argue that the thinness may reflect a 
young stellar population that is at least 500 Myr in age to account 
for the presence of RCGs. The coldness of the superthin component
may reflect young stars trapped in resonances at the bar ends. Such morphological features, called `ansae', are seen in external galaxies and simulations (\citealt{Martinez-ValpuestaKnapenButa2008}, \citealt{Athanassoula+2015}, \citealt{Athanassoula2016}). Complex structures like these may complicate the determination of the long bar length and, indeed, the projected properties here are not symmetric about the Galactic Centre, even accounting for the different distances \citep{WeggGerhard&Portail2015}. At the present time, it is not possible to determine a definitive stellar age for either component, which is clearly an important test. We may alternatively be observing the beginnings of loosely wound spiral arms emerging from the ends of the bar, which, as they twist into our line-of-sight, would account for an increasing density of young stars at both ends. The presence of a prominent star formation region at the receding end of the bar, and associated with the $Scutum$ arm, has been previously reported (\citealt{Lopez-Corredoira+1999}).

\section{Conclusion}\label{sec:conclusion}

In this paper we measured quantitative parameters of the Milky Way's (X/Peanut)--shaped structure from the Fourier $n=6$ component (cosine term, $B_6$) of its isophotes, extracted from 3.4\,$\mu$m and 4.6\,$\mu$m \WISE\ wide-field imaging. From the radial $B_6$ profile extracted with the $IRAF$ task \ifit, we determined the X/P length, height above the disc plane, as well as its orientation angle with respect to our line-of-sight to the Galactic centre. Specifically, we determined an intrinsic peanut radius of \Rpiint $= 1.67\pm0.27$ kpc, a height $z_{\it \Pi} = 0.65\pm0.17$ kpc, and a viewing angle of $\alpha = {37\degree}^{+7\degree}_{-10\degree}$. Using the X/P structure as a proxy of the Milky Way's long bar, we conclude that the latter is oriented at the same angle $\alpha$ and has an expected radius of $\approx 4.16\pm0.68$ kpc, but could possibly be as short as $3.24\pm0.54$ kpc. Our results are based on the picture in which the long bar and the elongated X/P structure of the Milky Way are not distinct and mis-aligned components, but are different regions of the same structure. Tilted at $\approx 37\degree$ from an end-on orientation, we find that this structure is viewed at a wider angle than conventionally thought for the triaxial `bulge' region ($\sim\!27\degree$) and a narrower angle than conventionally thought for the long thin bar ($\sim\!43\degree$).

The Milky Way appears to be a typical X/P galaxy, consistent with the CG16 scaling relations between the various X/P diagnostics (length, height and integrated strength of the peanut instability), as well as the observed correlation of $v/\sigma$ with peanut length and strength. The X/P strength parameter appears however to be marginally higher than the trend observed in nearby X/P galaxies, which is possibly a consequence of projection effects but may alternatively point to an enhancement in the Galaxy's X/P strength caused by accretion from its satellites. Additionally, we find tentative evidence of a $North-South$ asymmetry in the X/P feature, possibly reflecting the Galactic bar's past buckling phase that led to the formation of the peanut. We performed a photometric decomposition of the major axis surface brightness profile, in both \WISE\ bands, modelling the data with an exponential profile for the disc and Gaussian functions for the various spiral arms. We performed this in both the eastward and westward directions (with respect to the Galactic North) and obtained an average scale length of the disc of $h = 2.54\pm0.16$ kpc, in good agreement with the literature. As with other nearby X/P galaxies, the Milky way obeys the CG16 scaling relations when the peanut metrics are re-scaled by $h$, lending further support to the disc origin of the peanut (\citealt{Shen+2010}; \citealt{Ness+2012,Ness+2013}; \citealt{DiMatteo+2014, DiMatteo2016}).

\bibliographystyle{mn2e}
\bibliography{C.etal.2017}

\section{Acknowledgements}

We thank Dustin Lang for kindly providing the \UNWISE\ wide-field images of the Milky Way. AWG was supported under the Australian Research Council's funding scheme DP17012923. This publication makes use of data products from the Wide-field Infrared Survey Explorer, which is a joint project of the University of California, Los Angeles, and the Jet Propulsion Laboratory/California Institute of Technology, funded by the National Aeronautics and Space Administration.
\clearpage
\appendix

\setcounter{figure}{0}
\renewcommand{\thefigure}{A\arabic{figure}}

\section{Milky Way Photometric Decomposition}\label{sec:mwdecomp} 

\subsection{Integrated Light Approach}\label{sec:disc}

CG16 have shown that the X/P parameters of external galaxies are not arbitrarily distributed, but define specific scaling relations. The X/P length and height are correlated with each other, and both further correlate with the strength of the structure. Additionally, X/P galaxies also show a weak trend between their $v/\sigma$ ratio and the X/P length and strength. These trends hold when the various parameters are expressed either in kpc or in units of the host disc's scale length $h$. 

To investigate how the Milky Way fits into this picture, we determined its disc scale length by fitting its major axis surface brightness profile, i.e. the surface brightness as a function of galactic longitude $l$, in the mid-plane (galactic latitude $b=0$). This is similar to a typical galaxy decomposition, but it involves an extra step to correct for the fact that our vantage point is inside the galaxy being modelled. We first assume that the planar offset of the Sun is negligible, and that the disc (out to $\sim\! 8$ kpc) has an exponentially declining intensity profile given by:

\begin{equation}
I(r) = I_0 {\rm exp} (-r/h)
\label{equ:exp}
\end{equation} 

\noindent where $I_0$ is the intensity at the (Galactic) centre and $h$ is the exponential scale length of the disc. The galactocentric radial co-ordinate $r$ is expressed in heliocentric co-ordinates ($R,l,b$) as:

\begin{equation}
r (R,l;b\!=\!0) = \sqrt{R_0^{2} + R^2 -2RR_0\:{\rm cos}(l)} .
\label{equ:r}
\end{equation}

As we assume the Sun to be embedded in the disc plane, the observed intensity in a particular direction along the mid-plane (given by $l$ alone) is the integrated light from the position of the Sun to infinity:

\begin{equation}
I(l) = \int_0^{\infty} I(R',l;b\!=\!0) {\rm d}R' .
\label{equ:model}
\end{equation}

Assuming that the optical depth is also negligible (a reasonable assumption for our particular dataset), Equation \ref{equ:model} represents the model being fit to the observed mid-plane brightness profiles extracted from our wide-field imaging data, and corrected for dust absorption and IR glow (see \S \ref{sec:disc}). In the case of a single-component exponential model, $I(R',l,b\!=\!0)$ is simply given by Equation \ref{equ:exp}, with $r$ expressed as in Equation \ref{equ:r}. However, any azimuthally symmetric radial profile can be used, and in fact we employ additional components to capture the various spiral arms we observe in the data. 

\subsection{Disc Scale Length from \textbf{\textit{WISE}} Data}\label{sec:disc}

We obtained the scale length ($h$) of the Milky Way's disc from the photometric decomposition of its major axis surface brightness profile (SBP), correcting for the fact that we are observing the disc from within, as detailed in \S \ref{sec:disc}.

The surface brightness profiles were extracted by taking image ``cuts" along the disc mid-plane. While discs are generally approximated to have exponentially declining light profiles, in practice they often display complicating features such as spiral arms, which induce ``bumps" in the light profile. Because of the asymmetry induced by the Milky Way's various spiral arms, we again analysed the $E$ and $W$ sides separately. 

\begin{figure}
	\includegraphics[width=0.49\textwidth]{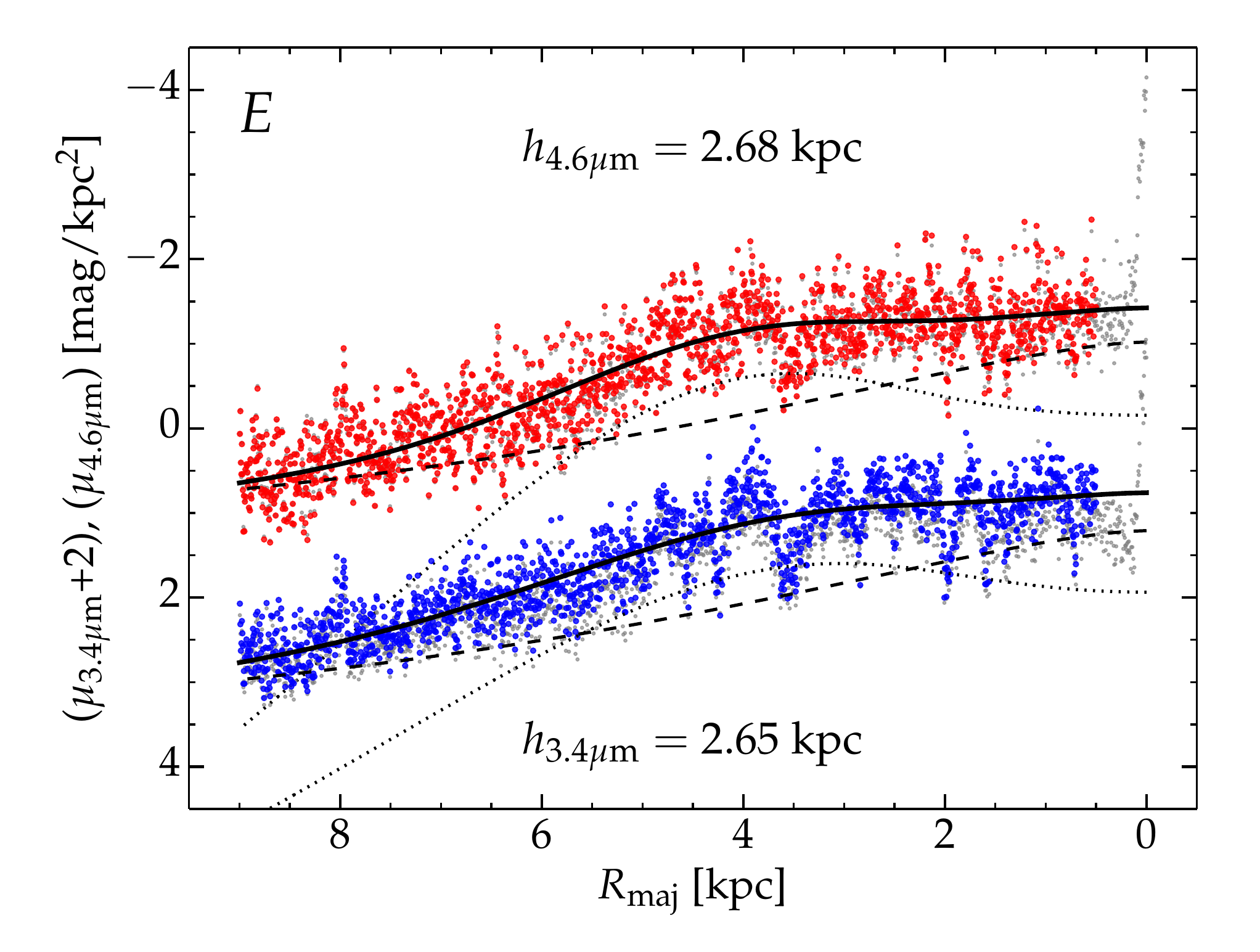}
	\caption{The eastward major axis surface brightness profile of the Milky Way at $3.4\mu$m (blue data) and $4.6\mu$m (red data). The models (black curves) consist of an exponential disc (dashed curves) and a Gaussian ring (dotted curves), the latter capturing the ($Scutum$ + $far~3~kpc$) spiral arms as single, `blended' features.} 
\label{fig:discfit-ring}
\end{figure}

\begin{figure*}
	\includegraphics[width=0.49\textwidth]{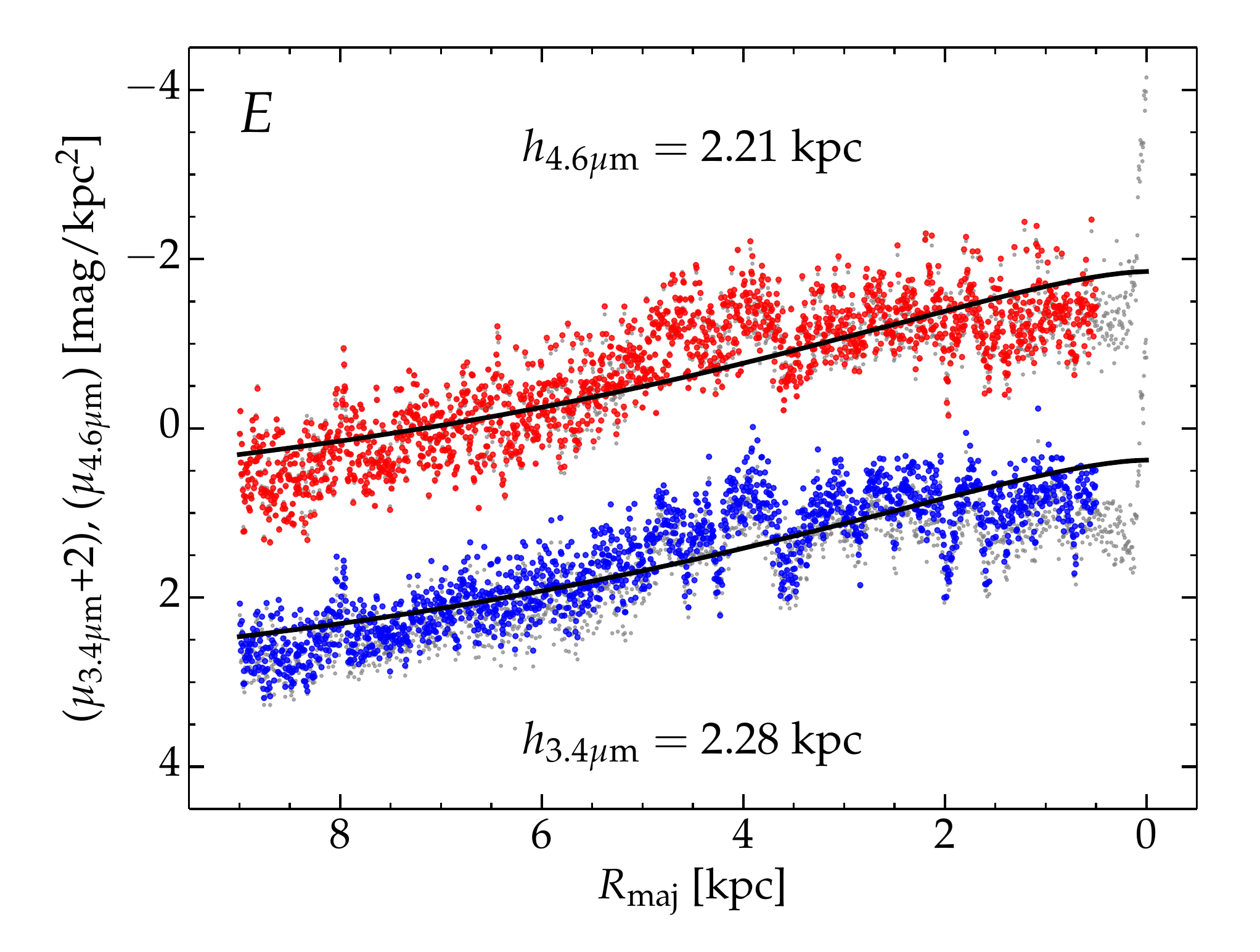}
	\includegraphics[width=0.49\textwidth]{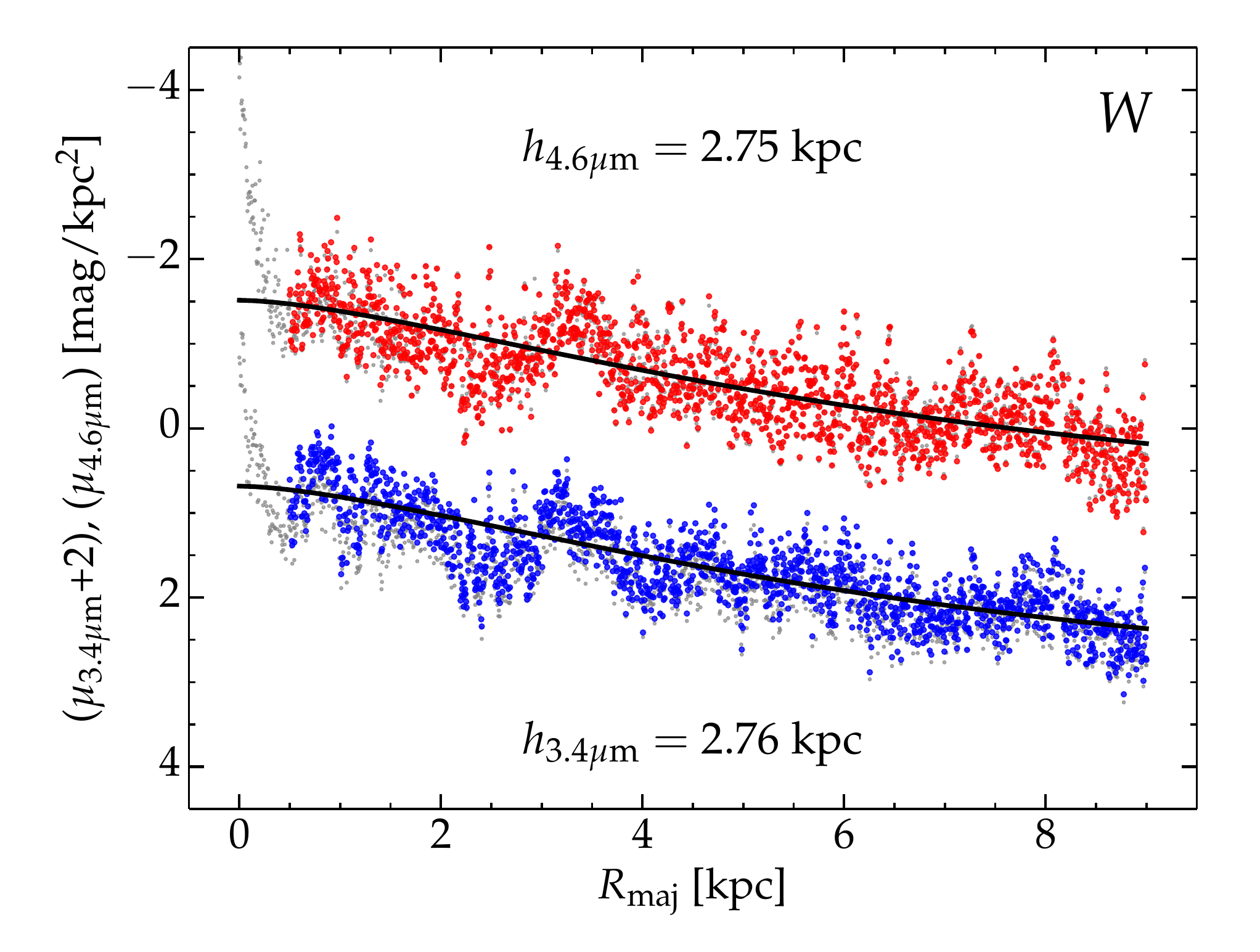}\\
	\includegraphics[width=0.49\textwidth]{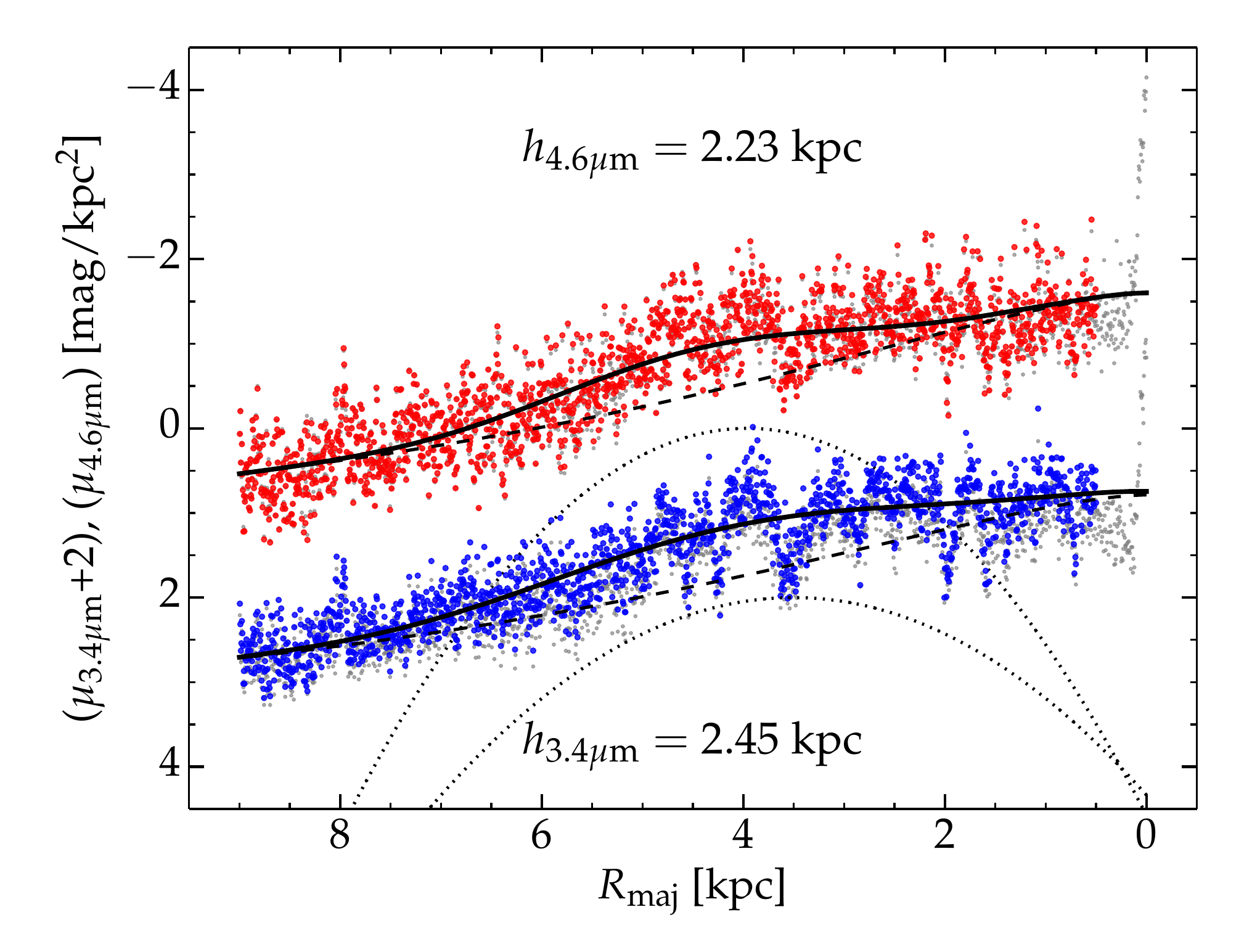}
	\includegraphics[width=0.49\textwidth]{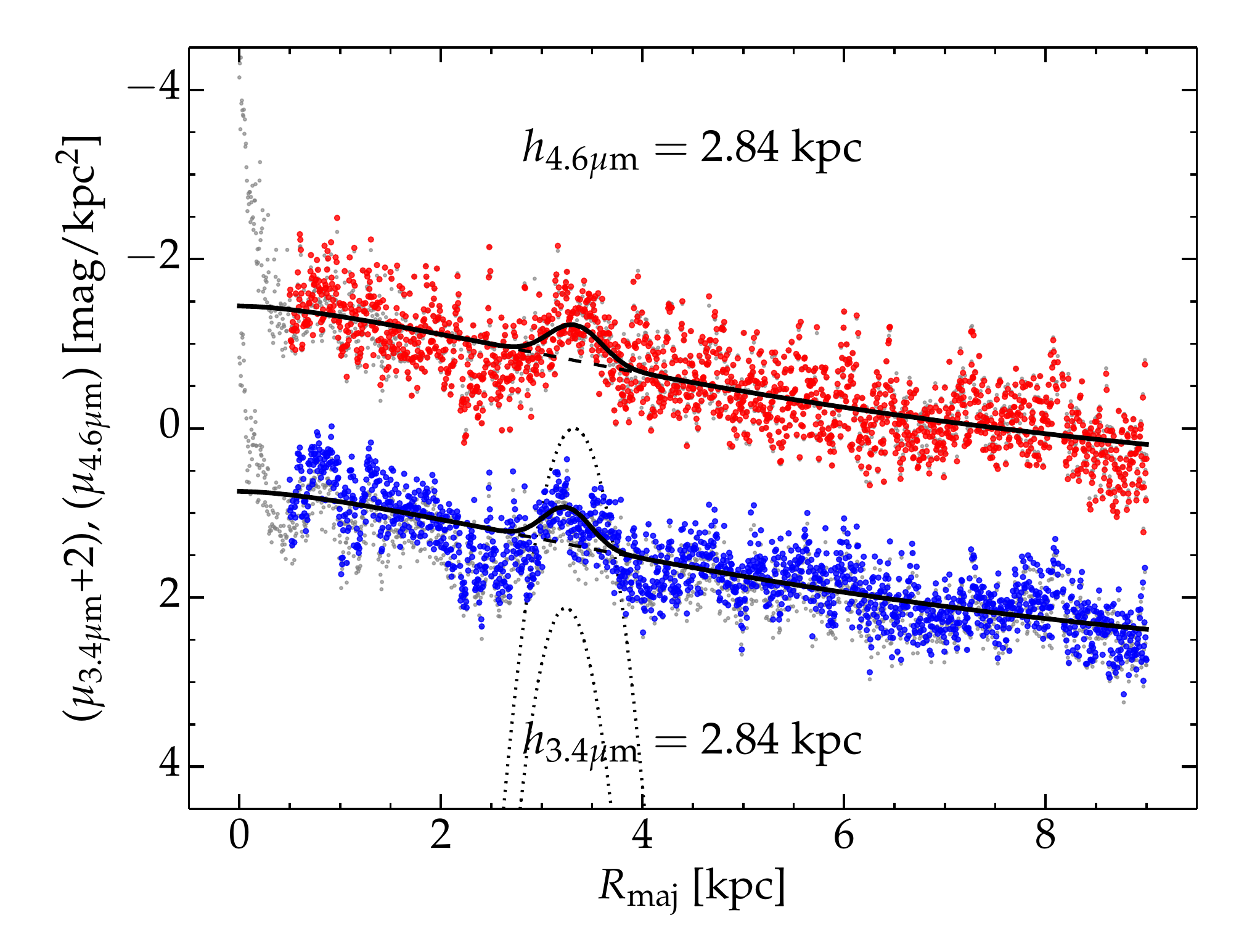}\\
	\caption{1D cuts in the plane of the disc to the $East$ of the Galactic Centre (left-hand side) and to the $West$ (right-hand side). Blue and red data correspond to the $3.4\mu$m and $4.6\mu$m images, while black curves represent the best-fitting model, corrected for our vantage point within the disc and assuming Sun's Galactocentric distance of 8.2 kpc. Insets indicate the best-fit disc scale length $h$ for each panel. {\bf Top:} Single exponential models. {\bf Bottom:} (exponential disc + 1 Gaussian spiral arm) models. {\bf Bottom:} (exponential disc + 2 Gaussian spiral arms). See main text for a discussion on individual spiral arms and their modelling.} 
\label{fig:discfit}
\end{figure*}

The raw major axis light profiles are shown in Figures \ref{fig:discfit-ring} and \ref{fig:discfit} through grey symbols. They were further corrected for the effects of dust, particularly dust glow and extinction. From \cite{Li&Draine2001} (see their Fig. 10) we estimated dust glow to be $\approx 1/13$ of the stellar emission at $3.4\mu$m and $\approx 1/8$ at $4.6\mu$m. We further estimated the dust absorption at these wavelengths from extinction in the $V-$band. From Tab. 3 of \cite{Nozawa&Fukugita2013} we used the ratios $A_{3.4\mu {\rm m}}/A_V = 0.0346$ and  $A_{4.6\mu {\rm m}}/A_V = 0.0201$. The major axis $A_V$ profile was extracted from the all-sky $A_V$ extinction maps of \cite{Rowles&Froebrich2009}, and is shown in Figure \ref{fig:AV_profile}. The dust-corrected surface brightness profiles are shown in Figures \ref{fig:discfit-ring} and \ref{fig:discfit} as blue symbols (3.4$\mu$m) and red symbols (4.6$\mu$m). As dust is typically more centrally concentrated in disc galaxies, the net effect of these corrections was to slightly steepen the SBPs compared to raw cuts.

\begin{figure}
	\includegraphics[width=0.49\textwidth]{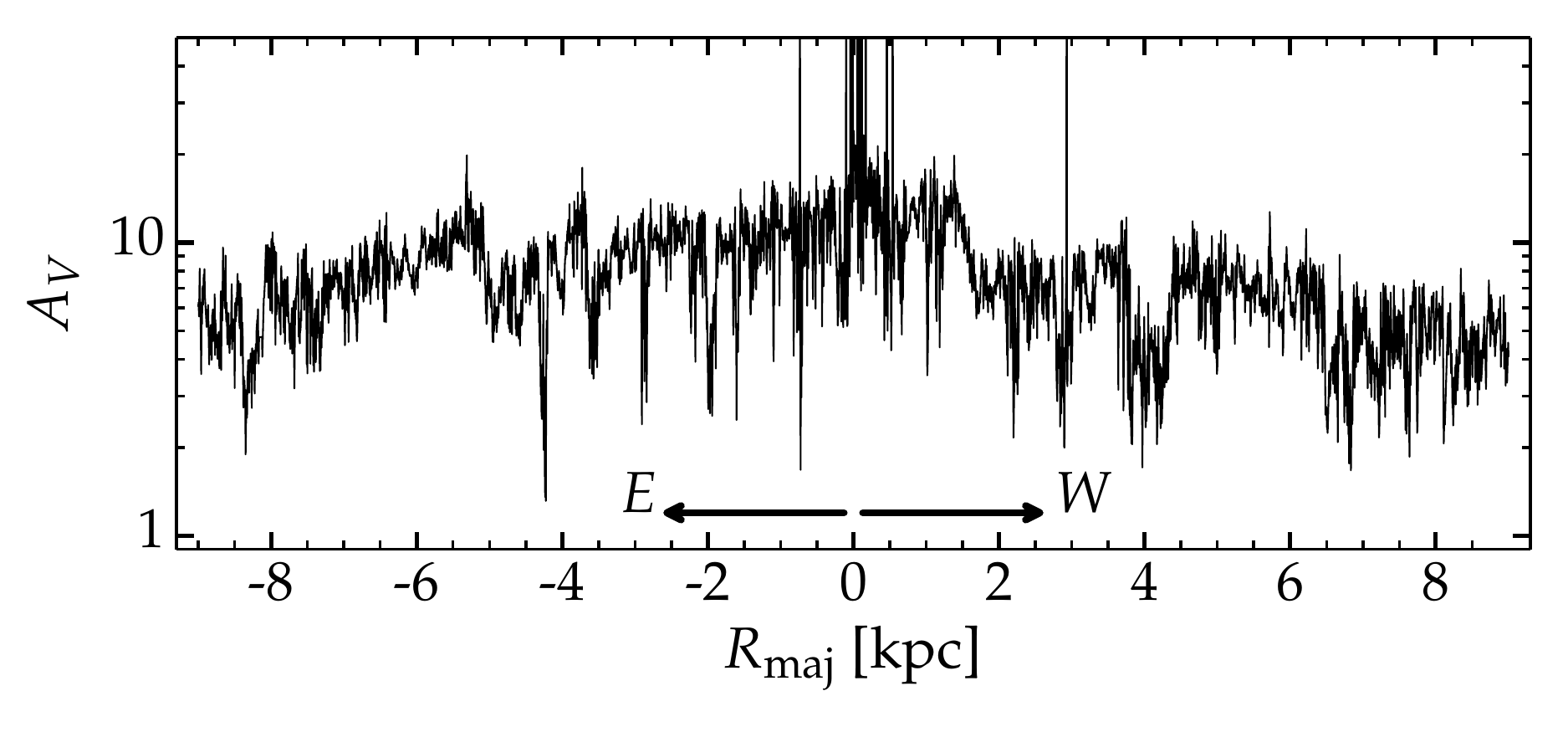}
	\caption{The $V$--band extinction profile along the major axis (disc mid-plane) extracted from the dust maps of \citealt{Rowles&Froebrich2009}.} 
\label{fig:AV_profile}
\end{figure}

While it is tempting to model spiral arms in the usual manner, as Gaussian rings, one must be mindful of the fact that they have a logarithmic nature, increasing their distance from the centre as they wind around azimuthally. We see this exemplified by the $Scutum$ arm, which peaks at different spatial scales in the two directions about the Galactic Centre, i.e. at $\sim 4.5$ kpc in the $E$ and at $\sim 8$ kpc in the $W$. We did nevertheless first attempt to model the arms as Gaussian rings, employing the same technique of integrating the light along lines of sight (\S \ref{sec:mwdecomp}). Thus, a Gaussian ring appears to take the form shown in Figure \ref{fig:discfit-ring} through the dotted curves. At the centre, the line-of-sight crosses perpendicular to the ring, so the SB value, given by twice the integral over the ring's thickness, is relatively low. By contrast, at the ring's radius, the line-of-sight is tangential to the ring, running $along$ it, so the integrated light reaches a maximum (bump) here, and gradually declines beyond this point. As noted above, a realistic spiral arm always has a lower curvature (or pitch angle) than a ring, which implies that at its tangent point, a line of sight runs a longer distance along the spiral arm than it would along a more curved ring. Therefore, the SB profile of a spiral arm has a stronger Gaussian-like bump and a weaker flattening central tail than a ring. After experimenting with both functions we found the pure Gaussian to give more robust and consistent results, and so chose this form for modelling the spiral arms. 

We modelled the data with increasing levels of sophistication. This is shown in Figure \ref{fig:discfit}, where the left-hand panels correspond to the eastward SBP while the right-hand panels to the westward SBPs. On the eastward side the data shows the $Scutum$ spiral arm as a rather prominent bump at $\sim 4.5$ kpc, as well as the less prominent {\it far 3 kiloparsec arm} as a feature centred at $\sim$ 3 kpc. The dip occurring at $\sim 3.5$ kpc is due to dust crossing the disc mid-plane, and is more pronounced (as expected) in the bluer filter. The westward SBPs show the {\it near 3 kiloparsec arm} at just beyond 3 kpc, and again the $Scutum$ (or $Scutum$-$Centaurus$) arm, this time at $\sim$ 8 kpc. We began by modelling the data on both sides with just an exponential profile (Figure \ref{fig:discfit} top panels). We further added a single spiral arm component (bottom panels) to the models, in each direction. Finally, we modelled both profiles with an exponential disc component and two spiral arm components, in each direction. We show these best-fit models in the main text of the paper, in Figure \ref{fig:discfit_best}.

We adopt a `global' value of the disc's scale length of $h$=2.54$\pm$0.16 kpc, the average of the best-fit (disc+2 spiral arms) models, in both filters and in the two directions. This result is in good agreement with the literature. For comparison, \cite{Licquia&Newman2016} report an average scale length, in the infrared, of $2.51^{+0.15}_{-0.13}$ kpc, from a Bayesian averaging method of literature measurements. We also refer the reader to \cite{Bland-Hawthorn&Gerhard2016} for a useful review on the Milky Way's structure. Finally, we note that a bar component, although faint, could also in principle be added to the models. We chose however not to include such a component since it is not well constrained by the data (which is additionally most affected by dust on the central spatial scales, where the bar is observed) and is thus degenerate with the spiral arm components.

\section{Derivation of the X/P Absolute Length and Viewing Angle}\label{sec:deriv}

\subsection{Derivation Based on Stewart's Theorem}

Equations \ref{equ:peanut-length} and \ref{equ:delta} in the main text, which yield the X/P length ($R_{\it \Pi}$) and viewing angle ($\alpha$), were derived by solving a system of two equations with the two quantities as the unknowns. The geometry of the problem is illustrated in Figure \ref{fig:geometry-ap}, which is analogous to Figure \ref{fig:geometry} but with different notation, to ensure clarity in this derivation. 

The first equation relating $R_{\it \Pi}$ and $\alpha$ came from considering the similar triangles $\bigtriangleup SAC$ and $\bigtriangleup SA'A''$. The fundamental theorem of similar triangles states that:

\begin{equation}
\frac{SA''}{SC} = \frac{A'A''}{AC}
\end{equation}

\noindent Analogously, from the similar triangles $\bigtriangleup SBC$ and $\bigtriangleup SB'B''$ it follows that:

\begin{equation}
\frac{SC}{SB''} = \frac{BC}{B'B''}
\end{equation}

\noindent As the two sides of the X/P structure are assumed to be equal ($A'C = B'C$), then $A'C\rm{sin}\alpha = B'C\rm{sin}\alpha = A'A'' = B'B''$, so, from B1 and B2, it follows that:

\begin{equation}
 \frac{AC\cdotp SA''}{SC} = \frac{BC\cdotp SB''}{SC}
\end{equation}

\noindent Making the substitutions $SA'' = SC - A'C\rm{cos}\alpha$ and $SB'' = SC + B'C\rm{cos}\alpha$, and simplifying the denominators, B3 becomes:

\begin{equation}
AC (SC - A'C\rm{cos}\alpha) = BC (SC + B'C\rm{cos}\alpha)
\end{equation}

\noindent Rearranging and using the notation of Figure \ref{fig:geometry}, we obtain the first equation which relates $R_{\it \Pi}$ and $\alpha$, namely:

\begin{equation}
{\rm cos}\alpha = \frac{R_0}{R_{\it \Pi}} \frac{R_\beta - R_\gamma}{R_\beta + R_\gamma} \equiv \eta \frac{R_0}{R_{\it \Pi}}
\end{equation}

\begin{figure}
\includegraphics[width=1.\columnwidth]{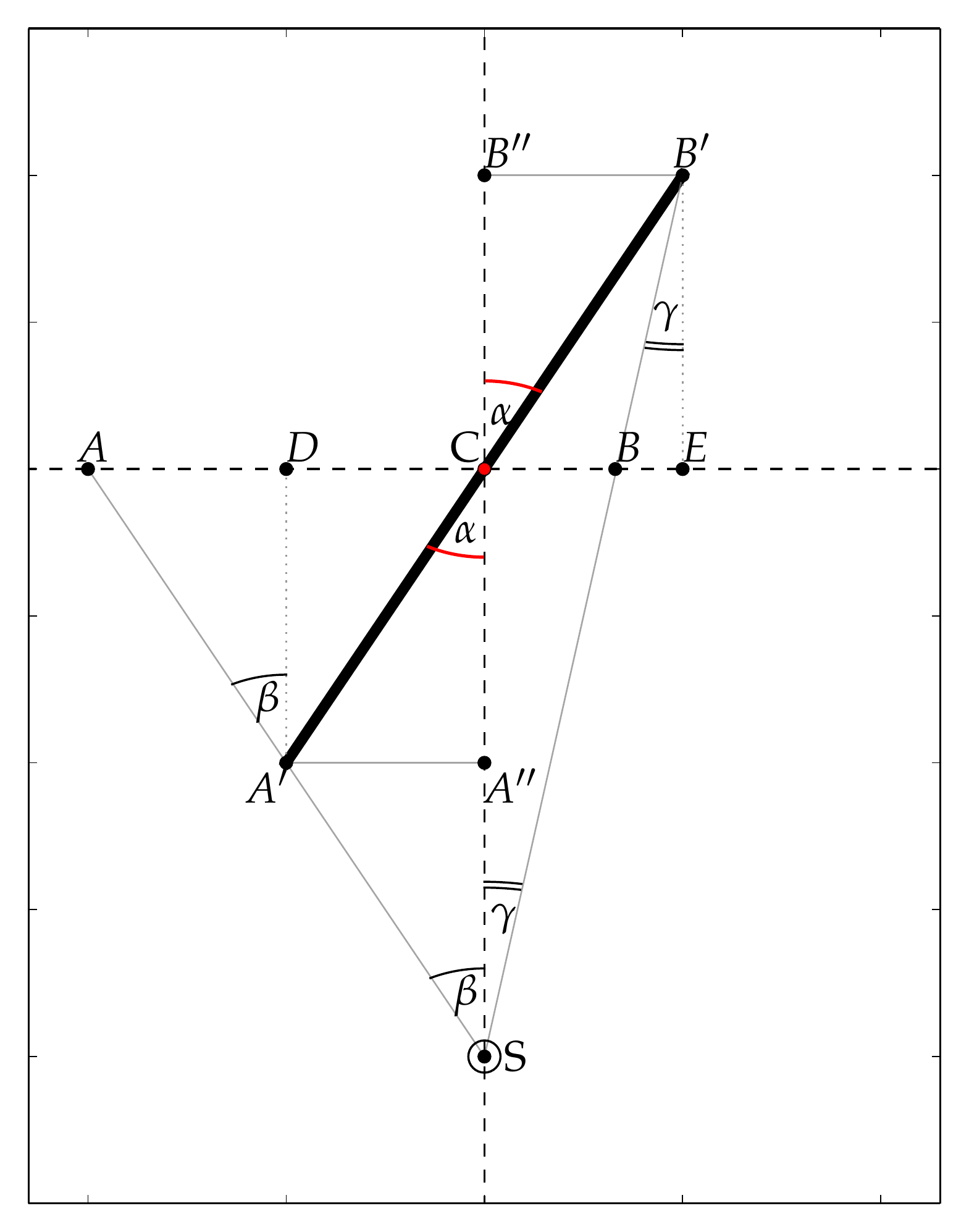}
\caption{Schematic of the (Sun+peanut) configuration, analogous to Figure \ref{fig:geometry} but with different notation used throughout the derivations in the Appendix. $S$ corresponds to the Sun, $C$ to the Galactic Centre and the thick line represents the X/P structure, orientated at a viewing angle $\alpha$.}
\label{fig:geometry-ap}
\end{figure}

The second equation relating $R_{\it \Pi}$ and $\alpha$ is obtained from Stewart's theorem. In particular, in $\bigtriangleup CAS$, with $CA'$ as the cevian, Stewart's theorem yields:

\begin{equation}
AC^2 \cdotp SA' + SC^2 \cdotp A'A = SA(A'C^2 + SA'\cdotp A'A) 
\end{equation}

\noindent where $SA = SC/$cos$\beta \equiv R_0/$cos$\beta$, and $SA'$ and $A'A$ can be obtained from the similar triangles $\bigtriangleup SA'A''$ and $\bigtriangleup SAC$, as follows:

\begin{equation}
\begin{split}
\frac{SA'}{SA} &= \frac{SA''}{SC} \Leftrightarrow  \frac{SA'{\rm cos}\beta}{R_0} = \frac{R_0 - R_{\it \Pi}{\rm cos}\alpha}{R_0} \Rightarrow \\
\Rightarrow SA' &= \frac{R_0 - R_{\it \Pi}{\rm cos}\alpha}{{\rm cos}\beta}
\end{split}
\end{equation}

\noindent and

\begin{equation}
\begin{split}
A'A &= SA - SA' \\
&= \frac{R_0}{{\rm cos}\beta} - \frac{R_0 - R_{\it \Pi}{\rm cos}\alpha}{{\rm cos}\beta }\\
&=\frac{R_{\it \Pi} \:{\rm cos}\alpha}{{\rm cos}\beta}
\end{split}
\end{equation}

\noindent Noting that $AC = R_{\beta}$ and using the expressions in B5, B7 and B8, equation B6 becomes:

\begin{equation}
\begin{split}
\frac{R_{\beta}^2 R_0 (1-\eta)}{{\rm cos}\beta} + \frac{R_0^3 \eta }{{\rm cos}\beta} = \\
\frac{R_0} {{\rm cos}\beta} \left[ R_{\it \Pi}^2 + \frac{R_0\eta (R_0 - R_0\eta)}{{\rm cos}^2\beta} \right] .
\end{split}
\end{equation}

\noindent Having substituted all (cos$\alpha$) terms through B5, the only unknown in B9 is $R_{\it \Pi}$, and re-arranging for it yields the required Equation \ref{equ:peanut-length}. The uncertainty in $R_{\it \Pi}$ is propagated from $\beta$, $\eta$ and $R_\beta$ and is given by:\\

\begin{equation}
\label{equ:err_radius}
\begin{split}
 \delta R_{\it \Pi} = &\frac{R_{\it \Pi}}{2}  \left\{ \left[2R_\beta (1-\eta) \delta R_\beta \right]^2 + \left[R_0^2 \left(1+\frac{2\eta-1}{{\rm cos}^2\beta - R_{\beta}^2}\right) \delta\eta \right]^2 +\right.\\
&+\left. \left( \frac{2\eta R_0^2{\rm sin}\beta \delta\beta}{{\rm cos}^3\beta}\right)^2 \right\} ^{1/2} ,
\end{split}
\end{equation}  

\noindent where $\delta\beta$ is the uncertainty in $\beta$, and $\delta R_\beta$ is obtained from $\delta R_\beta  = \sqrt{(R_0\delta\beta)^2 + (\beta \delta R_0)^2}$, which assumes the small angle approximation ${\rm tan}\beta \approx \beta$ and an uncertainty in $R_0$ of $\delta R_0$. In B10, $\delta\eta$ is the uncertainty in $\eta$, given by:

\begin{equation}
\delta\eta = \frac{2}{({\rm tan}\beta + {\rm tan}\gamma)^2}\sqrt{ [{\rm tan} \gamma\: \delta({\rm tan}\beta)]^2 +  [{\rm tan} \beta\: \delta({\rm tan}\gamma)]^2} ,
\end{equation}

\noindent which reduces, in the small angle approximation, to:

\begin{equation}
\delta\eta = \frac{2}{(\beta + \gamma)^2}\sqrt{ (\gamma\: \delta\beta)^2 +  (\beta\: \delta\gamma)^2}.
\end{equation}

\subsection{Viewing Angle and Uncertainties}

One can also first derive an expression for $\alpha$, and then recover $R_{\it \Pi}$, through B5. To do this we again start by defining two equations with the same two unknowns ($R_{\it \Pi}$ and $\alpha$). First, we see from Figure \ref{fig:geometry-ap} that:

\begin{equation}
AC = DC + AD = A'C\;{\rm sin}\alpha + A'D\;{\rm tan}\beta.
\end{equation}

\noindent Since $AC \equiv R_\beta$, $A'C \equiv R_{\it \Pi}$, and $A'D = A''C = R_{\it \Pi}\;{\rm cos}\alpha$, B13 can be re-written as:

\begin{equation}
R_\beta = R_{\it \Pi}\;{\rm sin}\alpha + R_{\it \Pi}\;{\rm cos}\alpha\;{\rm tan}\beta.
\end{equation}

\noindent Also from Figure \ref{fig:geometry-ap}, we see that:

\begin{equation}
\begin{split}
BC = EC - EB &= B'B'' - EB'\;{\rm tan}\gamma\\
			 &= B'C\;{\rm sin}\alpha - EB'\;{\rm tan}\gamma.
\end{split}
\end{equation}

\noindent But $BC \equiv R_\gamma$, $B'C \equiv R_{\it \Pi}$ and $B'E = B''C = R_{\it \Pi}\;{\rm cos}\alpha$, which, when substituted into B15, yields:

\begin{equation}
R_\gamma = R_{\it \Pi}\;{\rm sin}\alpha - R_{\it \Pi}\;{\rm cos}\alpha\;{\rm tan}\gamma.
\end{equation}
\noindent Dividing B14 and B16 by a factor of (cos$\alpha$) yields the equations:

\begin{equation}
\frac{R_\beta}{{\rm cos}\alpha} = R_{\it \Pi}({\rm tan}\alpha + {\rm tan}\beta),
\end{equation}

\noindent and

\begin{equation}
\frac{R_\gamma}{{\rm cos}\alpha} = R_{\it \Pi}({\rm tan}\alpha - {\rm tan}\gamma].
\end{equation}

\noindent Further dividing B17 by B18, and making the substitutions $R_\beta = R_0\;{\rm tan}\beta$ and $R_\gamma = R_0\;{\rm tan}\gamma$, results in:

\begin{equation}
\frac{R_0\;{\rm tan}\beta}{R_0\;{\rm tan}\gamma} = \frac{
R_{\it \Pi}9=({\rm tan}\alpha + {\rm tan}\beta}{R_{\it \Pi}({\rm tan}\alpha - {\rm tan}\gamma)},
\end{equation}
\noindent where $R_0$ and $R_{\it \Pi}$ simplify, and the equation rearranges into an expression for $\alpha$ as a function of only the two (measurable) angles $\beta$ and $\gamma$, which is:
\begin{equation}
\frac{2}{{\rm tan}\alpha} = \frac{1}{{\rm tan}\gamma} - \frac{1}{{\rm tan}\beta}.
\end{equation}

\noindent Having thus obtained the angle $\alpha$, one can the use it to calculate $R_{\it \Pi}$ through B5. The uncertainty in $\alpha$ can be computed by propagating the uncertainties in $\beta$ and $\gamma$. Since both angles are smaller than $\sim 10 \degree$, one can approximate tan$\beta \approx \beta$ and tan$\gamma \approx \gamma$. Equation B20 is re-written as:

\begin{equation}
{\rm tan}\alpha \approx \frac{2\beta\gamma}{\beta - \gamma} \equiv T .
\end{equation}

\noindent The uncertainty in $T$ is therefore:

\begin{equation}
\delta T = \frac{2}{(\beta - \gamma)^2} \sqrt{\gamma^4 \delta\beta^2 + \beta^4 \delta\gamma^2} ,
\end{equation}

\noindent which yields the upper and lower uncertainties in $\alpha$, namely $\delta^+\!\alpha$ and $\delta^-\!\alpha$ as follows:

\begin{equation}
\label{equ:err_delta}
\begin{split}
 \delta^+\!\alpha &= {\rm tan}^{-1}(T + \delta T) - {\rm tan}^{-1}(T) \\
& = {\rm tan}^{-1}(T + \delta T) - \alpha \\
 \delta^-\!\alpha &= {\rm tan}^{-1}(T) - {\rm tan}^{-1}(T - \delta T) \\
  &= \alpha - {\rm tan}^{-1}(T - \delta T) .
\end{split}
\end{equation}

\end{document}